\documentclass[reprint,superscriptaddress,aps,pra, superscriptaddress]{revtex4-2}
\usepackage[utf8]{inputenc}
\usepackage[T1]{fontenc}
\usepackage{bm}
\usepackage{natbib}
\usepackage{amsmath}
\usepackage{amssymb}
\usepackage{amsthm}
\usepackage{physics} 
\usepackage{nicefrac}
\usepackage{graphicx}
\usepackage{float}
\usepackage{url}

\renewcommand{\vec}[1]{\bm{\mathrm{#1}}}

\begin{document}
\title{Physical limits on electromagnetic response}
\author{Pengning Chao}
\author{Benjamin Strekha}
\author{Rodrick Kuate Defo}
\affiliation{Department of Electrical and Computer Engineering, Princeton 
University, Princeton, New Jersey 08544, USA}
\author{Sean Molesky}
\affiliation{Department of Engineering Physics, Polytechnique Montr\'{e}al,
Montr\'{e}al, Qu\'{e}bec H3T 1J4, CAN}
\author{Alejandro W. Rodriguez$^{1}$}
\email{arod@princeton.edu}
\begin{abstract}
 \noindent
 Photonic devices play an increasingly important role in advancing 
 physics and engineering, and while improvements in nanofabrication 
 and computational methods have driven dramatic progress in expanding 
 the range of achievable optical characteristics, they have also greatly 
 increased design complexity. These developments have led to heightened
 relevance for the study of fundamental limits on optical response. 
 Here, we review recent progress in our understanding
 of these limits with special focus on an emerging theoretical
 framework that combines computational optimization with conservation
 laws to yield physical limits capturing all relevant wave
 effects. Results pertaining to canonical electromagnetic problems
 such as thermal emission, scattering cross sections, Purcell
 enhancement, and power routing are presented. Finally, we identify
 areas for additional research, including conceptual extensions and
 efficient numerical schemes for handling large-scale problems.
\end{abstract}
\flushbottom
\maketitle
\thispagestyle{empty}


Photonics has become an indispensable tool of scientific discovery,
enabling key advances in communications~\cite{yariv2006photonics},
sensing~\cite{oh2021nanophotonic,zhang2021metasurfaces},
photovoltaics~\cite{garnett2020photonics},
computing~\cite{brunner2020photonics,shastri2021photonics}, quantum
engineering~\cite{dory2019inverse,chakravarthi2020inverse}, and many
other fields. At the center of the broad applicability of optical 
methods is a small but powerful set of design schemes for confining 
and transferring energy in time and space---notions such as index 
guiding, wave interference, polaritonics, and effective medium 
engineering---that provides physical intuition for extracting concrete 
functionality from the abstract mathematical richness within Maxwell’s 
equations. Each offers a mixture of distinctive capabilities and 
limitations, and determining the best approach or combination of 
approaches for any particular application remains a subject of 
continuing challenge for photonics design~\cite{liu2016fundamental}.

As a concrete example, consider the problem of enhancing light-matter
interactions via the photonic local density of
states (LDOS)---the Purcell effect---reducing to the
familiar Purcell factor $F_P = \frac{6}{\pi^2}
\left(\frac{\lambda}{2n}\right)^3 \frac{Q}{V}$ in the case where a 
single resonance dominates~\cite{joannopoulosphotonic}. Integrated
micro-resonators~\cite{vahala2003optical} based on index guiding can
achieve extremely long lifetimes (high quality factors $Q$) at the
expense of reduced spatial localization (large mode volumes
$V$). Electronic plasmon- and phonon-polariton resonances allow for tight
subwavelength confinement (small $V$) but suffer from high material
absorption (small $Q$) \cite{khurgin2015how}. Photonic
crystals~\cite{joannopoulosphotonic} and bandgap engineering provide a
flexible low-loss platform for manipulating light at the wavelength
scale but are limited in practice by achievable bandwidths and a
lack of forms exhibiting omnidirectional bandgaps. Metamaterials
offer conceptual simplicity in engineering exotic dispersions and
large LDOS~\cite{jacob2010engineering} but in practice are constrained
by fabrication limitations, the breakdown of effective-medium
approximations at large wavenumbers, and challenges related to light
coupling~\cite{sreekanth2014large,popov2016operator}.

Growing out of these general design principles, continued increases in
computational power have paved the way for the development of inverse
methods that, given a set of desired electromagnetic objectives and
constraints, exploit global~\cite{schneider2019benchmarking},
gradient-based~\cite{lalau-keraly2013adjoint,liang2013formulation,
christiansen2021inverse}, and data-driven optimization
algorithms~\cite{liu2018training,jiang2021deep} to search through
potentially millions of structural degrees of freedom in pursuit of 
ideal response characteristics. This capacity has greatly expanded the 
accessible design space, and has led to vast improvements in
device performance. 
However, for typical problems, the vast range 
of design possibilities and complicated interplay between standard 
objectives and constraints also makes it practically impossible to 
determine truly optimal structures, and one can at most hope for a
well performing local optimum~\footnote{Inverse methods may converge 
to structures that appear to reflect intuitive design principles, 
such as bowtie antenna and slot waveguide motifs for enhancing 
light-matter interaction~\cite{liang2013formulation,wang2018maximizing,
albrechtsen2021nanometer}.}. While our arsenal of design techniques 
and algorithms gives us enormous capability to tackle a wide range of 
engineering applications, it cannot rigorously answer a natural 
question of increasing relevance: what are the fundamental limits to 
optical control and how close are existing devices to reaching them? 

The notion of fundamental limits, encoded in bedrock principles like
the finite speed of light and second law of thermodynamics, is
ubiquitous in physics. Beyond added theoretical understanding, limits
have and continue to play an important role as signposts for further
technological improvement. Attempts to overcome the Abbe diffraction
limit
led in large part to the development of the field of super-resolution 
microscopy, with diverse techniques exploiting evanescent 
fields~\cite{betzig1986near,sanchez1999near-field,pendry2000negative},
nonlinear effects~\cite{sanchez1999near-field,
vicidomini2018sted}, and active temporal 
control~\cite{bates2013stochastic}. 
Knowledge of the physical origins of the factors forming the 
Shockley-Queisser limit~\cite{shockley1961detailed} for solar cell 
efficiency pointed the way to diverse developments in 
concentrators~\cite{lopez2007concentrator}, 
tandem~\cite{henry1980limiting}, and intermediate band 
photovoltaics~\cite{luque1997increasing}. 
The breakdown of familiar blackbody limits to nanoscale separations 
sparked interest in super-Planckian thermal 
devices~\cite{guo2012broadband,thompson2018hundred-fold}. 

\begin{figure*}[t!]
\centering 
  \includegraphics[width=2.0\columnwidth]{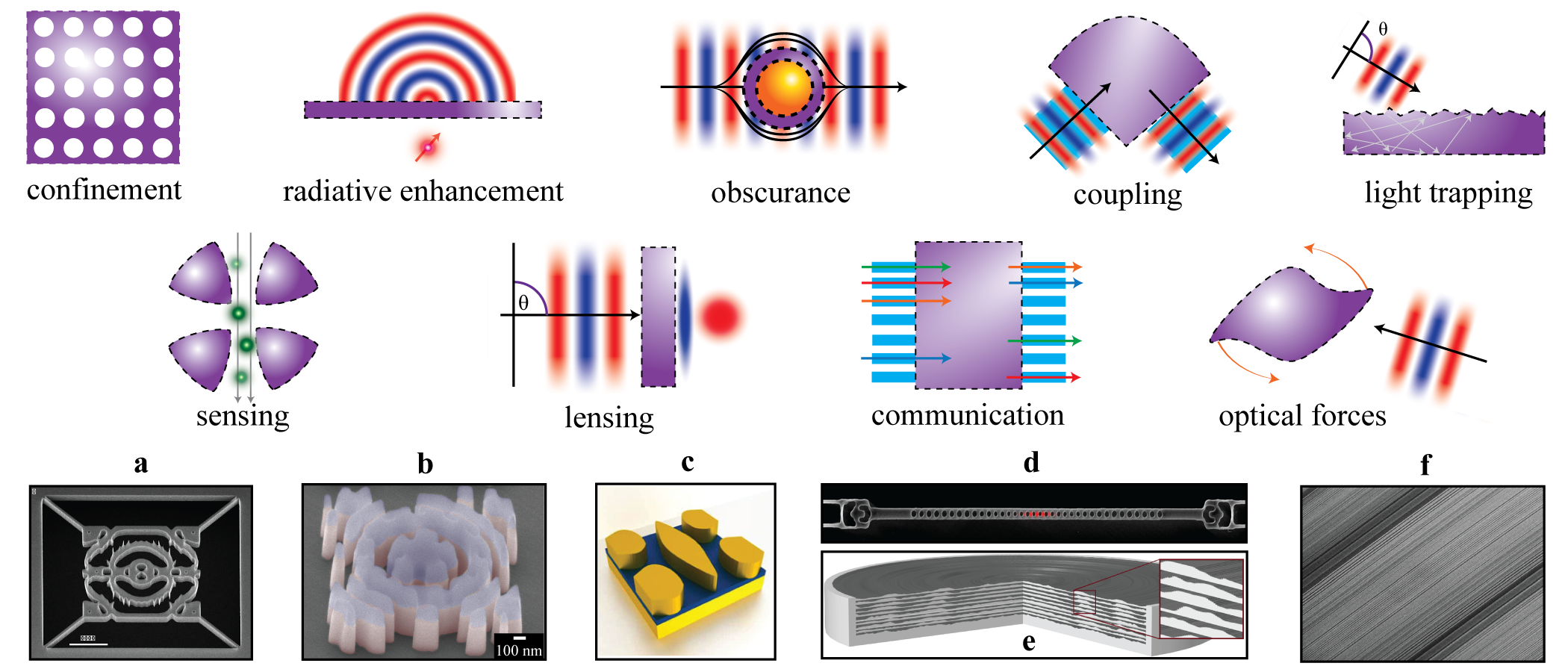} 
  \caption{\textbf{Applications of photonics and computational
  design.} The top part of the figure shows a collection of
  representative photonic functionalities. In each instance,
  performance is critically effected by the possibility of using
  (sub)wavelength-scale structuring to confine or transform optical
  fields and, as such, is intrinsically linked to physical limits on
  these phenomena. The bottom part of the figure depicts several
  (recent) examples of computationally synthesized devices for
  achieving improved performance in the operations shown above.
  Moving from left to right, these examples are \textbf{a} a
  dielectric cavity achieving nanometer scale confinement,
  Ref.~\cite{albrechtsen2021nanometer}; \textbf{b} a ``photon
  extractor'' enhancing the emission collection rate of a nitrogen
  vacancy center, Ref.~\cite{chakravarthi2020inverse}; \textbf{c} a
  frequency selective emitter for thermo-photovoltaic applications,
  Ref.~\cite{kudyshev2020machine}; \textbf{d} a photon-extractor
  (i.e. cavity and coupler) for diamond photonics,
  Ref.~\cite{dory2019inverse}; \textbf{e} and \textbf{f} large area,
  high-efficiency, metalenses, Ref.~\cite{christiansen2020fullwave}
  and Ref.~\cite{phan2019high} respectively. } \label{fig1_optProbs}
\end{figure*}

In this review, we first present a historical overview on the 
development of electromagnetic limits, following representative 
examples that illustrate a general thematic evolution essentially 
mirroring the history of optics itself: from simplifying and 
restrictive assumptions (homogeneity, quasistatics, and ray 
optics, etc.) pertinent to low-dimensional, deeply subwavelength, 
and large-etalon systems, toward increasingly sophisticated wave 
arguments applicable to any length scale. We then focus our 
discussion on an emerging general methodology for evaluating 
photonic design bounds based on optimization theory and conservation 
principles that follow either directly from Maxwell's equations 
or the identities of scattering theory. Originally developed as an 
instrument for investigating maximal 
scattering cross section limits~\cite{molesky2020fundamental,
gustafsson2020upper,kuang2020maximal}, the framework is applicable 
to a broad range of design problems where the objective can be 
expressed as a quadratic function of the 
fields~\cite{molesky2020hierarchical,kuang2020computational,
molesky2021comm}.
The constraints have clear physical meaning, limiting both the 
amplitude of the polarization response, important to power transfer, 
and the extent that the phase of the polarization response can be 
modified, with consequences on the engineering of resonances. 
To better handle problems involving near field effects and rapidly 
varying length scales, spatially localized constraints can be 
introduced, with a denser distribution of local constraints giving 
tighter bounds at the expense of higher computational 
complexity~\cite{molesky2020hierarchical,kuang2020computational}. 
In this sense the framework emphasizes the complementary role of limits 
and structural optimization: structural optimization enforces Maxwell’s 
equations exactly (up to computational discretization) and produces 
specific devices corresponding to local optima; the limits framework 
instead produces bounds that apply to all possible structures via 
conservation-law based constraints over spatial regions that can be 
viewed as a relaxation of Maxwell's equations. 

Through instructive results concerning thermal emission,
scattering/absorption cross sections, LDOS enhancement, and power
splitting, we describe physical implications behind various components
of the framework, drawing connections to well-known prior results and
demonstrating its broad applicability. For readers interested in the
mathematical details of the underlying optimization theory, we also 
recommend the excellent review by Angeris et al. 
Ref.~\cite{angeris2021heuristic}.
Finally, we detail remaining challenges and opportunities, including
the need for numerical methods that can handle larger systems,
generalizations to other physical settings beyond photonics, and
potential improvements to structural optimizations that may arise from
knowledge of optimal fields.

\section*{Historical overview}
Since the first measurement of the speed of light in vacuum by R\o mer, 
and the subsequent postulates of special relativity demanding that 
information cannot travel faster than $c$, rigorous proofs of 
subluminal energy velocity $v_e \leq c$ have been deduced from 
increasingly general assumptions, moving from homogeneous non-absorbing media, 
through the inclusion of dispersion, anisotropy, and 
nonlocality~\cite{brillouin2013wave,schulz1969energy,
loudon1970propagation,yaghjian2007internal}, to the simple unifying 
requirement of \emph{passivity}: materials which do no net work on 
electromagnetic fields~\cite{glasgow2001poynting,yaghjian2007internal,
welters2014speed}.
In complement, there has also been great interest in establishing
limits on minimal energy velocity or “slow light”, i.e., engineered
devices such as optical delay lines and
buffers~\cite{tucker2005slow-light,liu2001observation,hau1999light,
yariv1999coupled-resonator,soljacic2002photonic-crystal}. Under the
approximation of zero bandwidth, the delay experienced by a light pulse can
essentially be made indefinitely long, e.g., near the band edge of a
photonic crystal where the group velocity
vanishes~\cite{povinelli2005slow-light}. For finite bandwidths,
however, delay-bandwidth product limits (e.g., the uncertainty
principle) must set a fundamental lower bound. For a slow light
waveguide with an idealized linear dispersion relation across the
bandwidth of interest, the delay bandwidth product limit takes the
form $\Delta T \Delta f \leq \frac{L}{\lambda_c}(n_{avg}-n_{min})$,
where L is the length of the device, $\lambda_c$ is the free-space
bandwidth at the band center, and $n_{avg}$, $n_{min}$ are the average
and minimum effective index of refraction across the bandwidth,
respectively~\cite{tucker2005slow-light}. For the simple
case of explicitly 1D wave propagation and material structuring, more 
general bounds can be derived without the notion of an effective 
index of refraction and the assumption of idealized linear dispersion: 
$\Delta T \Delta f \leq \frac{1}{2\sqrt{3}}\frac{L}{\lambda_c}
\frac{\max\lVert\epsilon(z,f)-\epsilon_{ro}\rVert}{\epsilon_{ro}}$,
where $\epsilon(z,f)$ is the (possibly complex) relative dielectric
constant as a function of the propagation coordinate $z$ and frequency
$f$, $\epsilon_{ro}$ is the relative dielectric constant of the
background medium, and the maximum is taken across space and
bandwidth~\cite{miller2007fundamental,miller2007slow}. Further work 
is needed to extend these results to accommodate rigorous
full-wave descriptions of three-dimensional structures.

Related bandwidth arguments have also been used to derive performance bounds 
on optical cloaks---devices that eliminate or greatly reduce scattering
of incident light~\cite{fleury2015invisibility}. 
Most commonly, metamaterial cloaks are designed via transformation 
optics to carry out prescribed phase and amplitude manipulations 
by mapping coordinate transformations, $\mathbf{x} \to \mathcal{J} 
\mathbf{x}'$, onto effective homogenized
susceptibility parameters, $\{\epsilon, \mu\} \to
\{\epsilon', \mu'\} = \frac{\mathcal{J}^\dagger \{\epsilon, \mu\}
\mathcal{J}}{\mathrm{det} \mathcal{J}}$, in a design volume 
surrounding the object. 
Because light must go around the cloaked object and maintain an unperturbed wavefront,
the phase velocity in any cloak must be superluminal, necessitating 
the existence of dispersion in order to respect causality, and
thereby precluding perfect cloaking in vacuum over any finite
bandwidth~\cite{pendry2006controlling}. 
One way around this limitation is to consider ``ground-plane'' 
cloaking, wherein the object to be concealed is positioned adjacent to 
a reflecting boundary. 
With this shift in perspective, the original causality constraint 
does not apply, as the reflected waves from the cloak travel a 
shorter distance than reflected waves from the ground plane itself. 
Yet, because the wavefront within the cloak must now be delayed, 
the basic challenge posed by the need to respect delay-bandwidth 
limits remains, and, in the simplifying case of 1D waves, leads to 
a bound on the minimum thickness of the cloak $d \gtrsim
\frac{h}{n(n-1)}\frac{\Delta\omega}{\omega}$ as a function of the object size 
$h$ and the bandwidth $\Delta \omega$. 
Generalizing to 3D \cite{hashemi2011general}, any ground plane 
cloak must obey the inequality $V_c \geq V_{c}' / B$ where $V_c$ is the
volume of the cloak, $V_{c}'$ is combined volume of cloak and object,
$B$ represents the maximum achievable index contrast (eigenvalues of
$\mathcal{J}^T \mathcal{J}$). 
While instructive, determining attainable $\mathcal{J}$, and hence 
$B$, requires specific geometric analyses.

\begin{figure*}[t!]
\centering
\includegraphics[width=1.8\columnwidth]{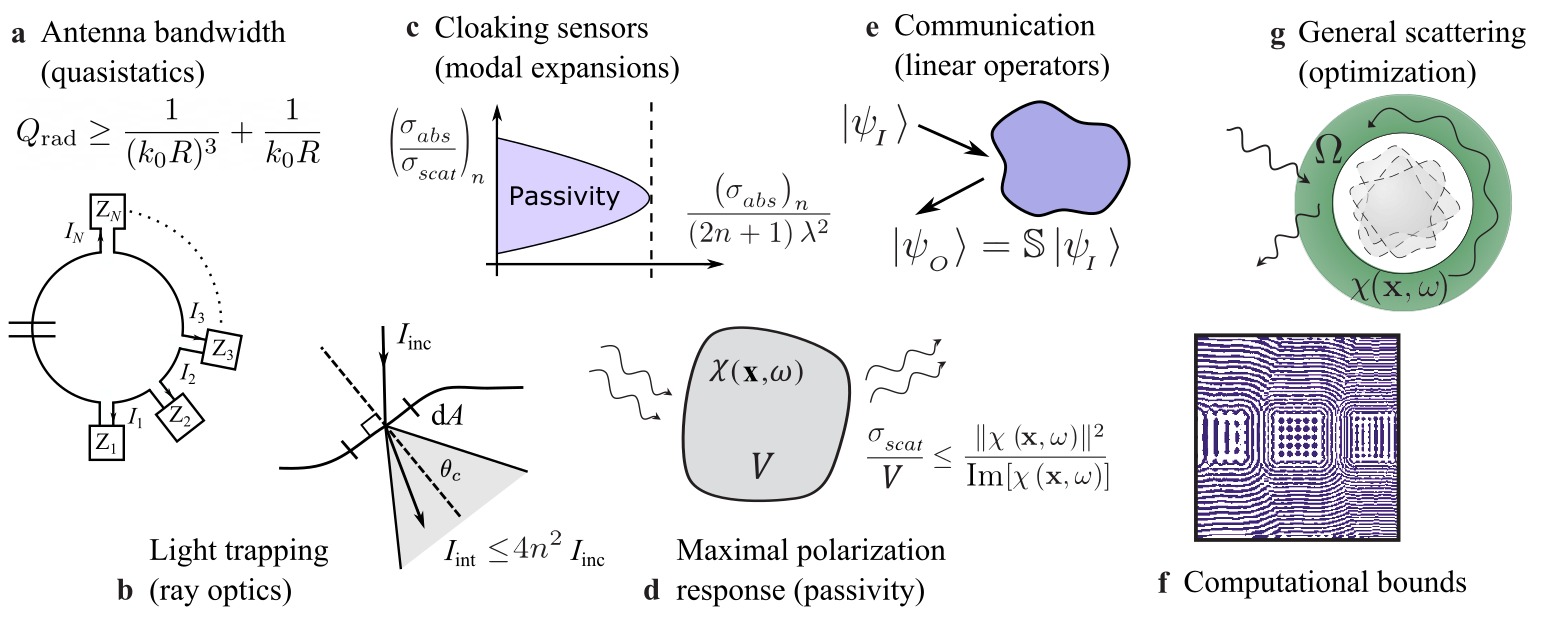}
\caption{\textbf{Overview of representative electromagnetic limits.} 
The figure illustrates a collection of representative photonic bounds,
arranged from left to right in roughly chronological order.
\textbf{a} Equivalent circuit of an omnidirectional antenna, 
adapted from \cite{chu1948physical}. For electrically small antennas, 
taking the quasistatic limit yields the celebrated Chu limit. 
\textbf{b} Yablonovitch limits to light trapping at a rough interface, 
combining geometric (ray) optics with a statistical description of the 
scattering angle~\cite{yablonovitch1982statistical}. 
\textbf{c} and \textbf{d} Limits based on passivity: for a passive 
device both the absorbed and scattered power must be non-negative. 
\textbf{c} shows the feasible region for cloaking as a function of 
absorption efficiency $\sigma_{sca}/\sigma_{abs}$ and net absorption
$\sigma_{abs}$ \cite{fleury2014physical}.
\textbf{d} illustrates upper bounds on scattering cross sections 
dependent on the material susceptibility $\chi$.
\textbf{e} Schematic showing a scattering matrix description of an 
optical device connecting input waves to output waves. Tools of
linear algebra such as singular value decompositions can be used to
analyze limits on communication and power transfer using
lightwaves~\cite{miller2000communicating,miller2007fundamental,
miller2017universal}.
\textbf{f} Suggested initial design for a multimode 2D Helmholtz 
resonator, based on the solution of the Lagrange dual problem of the design 
optimization~\cite{angeris2019computational}. 
\textbf{g} Schematic (adapted from \cite{jelinek2021fundamental}) 
representing a Lagrange dual framework for evaluating general photonic 
bounds using only knowledge of the design region $\Omega$ and material 
susceptibility $\chi$, through constraints based on conservation laws 
derived from Maxwell's equations. Overall we see a trend from early 
results that are problem or regime specific towards more recent limits 
with broader applicability.}
\end{figure*}

The intertwined concepts of optical delay and waveguide propagation
naturally lead to questions concerning optical
confinement~\cite{vahala2003optical}. Two main techniques are
commonly used to confine light, without relying on coupling to
material resonances~\cite{maier2007plasmonics}: index guiding, a
generalization of total internal reflection to wavelength-scale
cavities (e.g. whispering gallery or ring resonators), and bandgap
confinement, a generalization of Bragg scattering with embodiments in
photonic crystal waveguides and holey
fibers~\cite{joannopoulosphotonic}. Except for the simplest cases,
determining the propagation characteristics of a specific design
requires numerical computation, making sufficient conditions for the
existence of guided modes, in analogy with related variational
conditions for bounds states in quantum mechanics, conceptually and
practically useful. For instance, the displacement field $D_c$ of the
fundamental mode of a waveguide with permittivity profile $\epsilon$
and cladding profile $\epsilon_c$ was recently shown to necessarily
satisfy $\int D_c^* (\epsilon^{-1} - \epsilon_c^{-1}) D_c < 0$ within
the cladding~\cite{lee2008rigorous}. Relatedly, the degree of
localization achievable by a photonic crystal defect mode is
intuitively proportional to bandgap size~\cite{joannopoulosphotonic},
leading to recent variational bounds on the minimum index contrast
required to engineer 2D bandgaps~\cite{rechtsman2009method}.
Generalizations to incorporate conditions for dual-polarization and 3D
localization, along with considerations of
quasicrystalline~\cite{vardeny2013optics} and disordered
media~\cite{yu2021engineered} remain open problems.

Light confinement is also an integral tool for enhancing light--matter
interactions in optical modulators, lasers, and quantum devices. 
As detailed in the Introduction, a resonant mode enhances the power 
radiated by a nearby dipolar emitter in accordance with the
Purcell factor, which grows proportionally with the quality factor $Q$ (longer
lifetimes) and inversely with mode volume $V$ (higher field
intensities). Beyond modal descriptions that do not readily
generalize to multi-resonant systems~\cite{pick2017general}, or that
refer to specific geometric designs, the electromagnetic local density
of states (LDOS) stands as a fundamental figure of merit quantifying
optical response in arbitrary settings. By enforcing passivity for
scattered power~\cite{miller2016fundamental}, recent quadratic
optimization arguments have made it possible to constrain the
magnitude of achievable polarization response independent of
geometric or modal considerations. For a dipole emitter a small
distance $d$ away from a device enclosed within a half space, such
conservation arguments yield a material bound on the dominant
contribution of evanescent fields to LDOS enhancement at a single
wavelength, $\frac{\rho(\omega)}{\rho_0(\omega)} \leq \frac{1}{8\pi^3}
(\frac{\lambda}{d})^3
\frac{\lVert\chi(\omega)\rVert^2}{\Im \chi(\omega)}$, where $\rho_0(\omega)$ 
is the free-space LDOS. 
As detailed later in this Review, the positivity of scattered power alone cannot 
fully account for wave effects and resonance conditions, and in fact is 
only one of several other important constraints that limit optical 
response beyond quasi-static settings~\cite{molesky2020fundamental}. 
Passivity arguments based on Kramers--Kronig conditions 
similarly place limits on the frequency-integrated
material response of a medium, yielding sum rules of the form, 
$\int_0^{\infty} d\omega\,\frac{\rho(\vb{x}, \omega)
- \rho_0(\omega)}{\rho_0(\omega)} = 0$~\cite{barnett1996sum,scheel2008sum}. 
Both these arguments and related Thomas--Reich--Kuhn sum rules have 
in turn been used to derive upper bounds on nonlinear response, e.g., 
molecular hyperpolarizabilities~\cite{kuzyk2000physical,kuzyk2013sum}.

Another important aspect of localization relates to the focusing of
optical power from a source to a receiver. When restricted to systems
described by geometric optics, the conservation of
etendue~\cite{markvart2007thermodynamics} places a lower bound on how
tight the light rays from a source can be focused
down~\cite{ries1982thermodynamic}; accounting for wave effects,
scattering concentration bounds over input-output modal channels of
the form $\langle\lVert c_{out, \hat{u}}\rVert^2\rangle \leq \text{maxeig}
(\rho_{in})$ yield the maximum concentration of power achievable for
any linear combination of output channels represented by the unit
vector $\hat{u}$ in terms of the largest eigenvalue of $\rho_{in}$,
the density matrix describing the power flow and coherence across
input channels~\cite{zhang2019scattering}. Achromatic
metalenses~\cite{chung2020high,banerji2019imaging,
chen2018broadband,lin2019overlapping,shrestha2018broadband,
wang2018broadband} that focus several incident beams onto the same
focal spot are further restricted by causality and, thus,
delay-bandwidth limitations~\cite{presutti2020focusing}.

More broadly, limits on focusing are connected to the general theme of using 
light as a conduit for information and energy transfer. An early
constraint related to energy transfer is Kirchhoff’s law, equating the
emissivity and absorptivity of an object, often associated with the
second law of thermodynamics (detailed balance) and originally derived
under assumptions of geometrical optics and
reciprocity~\cite{kirchhoff1978verhaltnis,onnes1914simplyfied,robitaille2009kirchhoff}.
Generalizations of this concept via the formalism of a linear
“mode-converter” have been used as models of communication capacity,
and are in principle capable of accounting for wave effects and
non-reciprocal media~\cite{miller2017universal}. In particular,
information encoded in waves transferred between a source $S$ and
receiver region $R$, in free space, can be quantified via a Frobenius
norm, $\int_{V_S} \int_{V_R}
\lVert \mathbb{G}_{0}(\mathbf{x},\mathbf{x}')\rVert^2$, of the vacuum Green's
function connecting the two enclosing volumes, $V_S$ and $V_R$.
Adaptations to describe communication mediated by devices (e.g.,
lenses and multiplexers), which can strongly modify electromagnetic
fields and thus ``channel capacity'', remain an active area of
investigation~\cite{ellis2017performance,molesky2021comm}.

A related perspective on communication can be gleaned by considering
heat as a stochastic source of energy transfer. 
The blackbody limit as applied to radiative heat transfer constrains 
the flux emitted by a macroscopic body of area $A$ and temperature $T$ 
to be $H \leq \sigma T^4 A$, where $\sigma$ is the famous 
Stefan--Boltzmann constant~\cite{onnes1914simplyfied}. 
However, this result is only applicable to objects where all 
characteristic lengths are substantially larger than the thermal 
wavelength $\lambda_T=\frac{2\pi c 
\hslash}{k_B T}$, and hence does not explain, for instance, the power 
exchanged between two bodies held at different temperatures separated
by a subwavelength vacuum gap $d$; nor does it account for the
material constraints subsumed in the assumption of perfect absorption,
i.e. the difficulty of engineering absorption over a wide bandwidth in a
device of a limited size~\cite{mizuno2009black,
yoon2015broadband,magdi2017broadband}. Just as with LDOS, the
positivity of scattered power sets material constraints on the
achievable polarization response that waves originating in one body
may excite in another~\cite{miller2016fundamental}, yielding an upper
bound on the mutual absorption of light $\propto
\frac{\lVert\chi_S\rVert^2}{\mathrm{Im}\chi_S}
\frac{\lVert\chi_R\rVert^2}{\mathrm{Im}\chi_R} \int_{V_S} \int_{V_R}
\lVert \mathbb{G}_{0}(\mathbf{x},\mathbf{x}')\rVert^2$ (generalizing the
aforementioned communication bounds to incorporate material
considerations in the source/receivers), and a corresponding upper
bound on the spectrally integrated heat transfer of $H \leq \sigma T^4 A
\frac{2}{7(kd)^2} \frac{\lVert \chi\rVert^3}{\Im \chi}$. 
Going further, accounting for radiative losses due to mutual
scattering between bodies reveals the much tighter bound of $H \leq
2 \sigma T^4 A \big(\frac{\lambda_T}{d}\big)^2
\frac{\Im\chi}{\lVert \chi \rVert^2} \ln\left[\frac{\lVert \chi\rVert^2}{4 \Im \chi} \right]$. 
The origin of this reduced material scaling lies in the infeasibility
of achieving resonant optical response for all waves and is
elaborated on further below~\cite{venkataram2020fundamental}.

The difficulty in engineering blackbody response is directly related to limits
on the absorption of incident radiation, also known as light trapping in the context
of photovoltaic applications.
The celebrated Yablonovitch limit~\cite{yablonovitch1982statistical}, 
originally derived via a statistical description of rays scattering off 
rough surfaces, posits a maximum absorption enhancement factor 
$F \leq 4n^2$, compared to the expected single pass absorption 
$\alpha d$ of a weakly absorbing bulk film of thickness $d$ and 
absorption coefficient $\alpha$. 
The dependence on the refractive index $n$ enters via the total internal
reflection angle $\theta_c = \arcsin(1/n)$, which sets the emission
cone from which light can escape. 
Analyses of maximum absorptivities for films of thickness 
$d \lesssim \lambda$ have been carried out through modal 
decomposition techniques~\cite{yu2010fundamental},
allowing the associated limit to be expressed as a
$\frac{1}{\Delta\omega} \sum_m \sigma_{m,max}$, with the maximum
spectral absorption cross section for each mode $\sigma_{m,max}$
determined by specific material and geometric considerations. 
In the simplifying regime of a thin film supporting a single guided 
band for each polarization, this approach gives a limit absorption
enhancement of $F \leq \big(\frac{\lambda}{d}\frac{\alpha_{wg}}{\alpha} 
\frac{1}{2 n_{wg}}\big) 4 n_{wg}^2$, where $n_{wg}$ is the group index of 
the mode(s) and $\frac{\alpha_{wg}}{\alpha}$ characterizes the spatial
overlap between the mode profile and absorption layer. As examined
below, aside from their practical utility in predicting performance
for specific geometries, such modal summations can be employed to gain
a qualitative understanding of achievable absorption characteristics.
In contrast, limits based on maximal material response of the form
$\frac{\sigma_\mathrm{abs}}{V}
\leq \frac{\lVert\chi\rVert^2}{\Im\chi}$~\cite{miller2016fundamental}
remedy the need of geometric specificity (beyond a linear 
volumetric $V$ dependence), but can be shown to be loose 
beyond quasistatic settings, or in cases where it is not possible to 
achieve resonant response.

Besides transferring energy, light also imparts a
force~\cite{dienerowitz2008optical,macchi2009light}: the elastic
scattering of impinging photons on a body of size much larger than
$\lambda$ transfers a momentum of $\Delta p = 2 hf/c$. For bodies
with wavelength-scale features, the impact of wave effects and
nanostructuring on the scattering cross section becomes pronounced.
For quantum and thermal waves originating within bodies---often associated with Van der
Waals and Casimir forces---the situation is even more complicated due to the broadband and incoherent nature
of thermodynamic fluctuations. Despite these challenges, a
no-go theorem establishing the impossibility of repulsive interactions
between mirror-symmetric bodies separated through
vacuum~\cite{kenneth2006opposites} exists, as do recent bounds on
Casimir--Polder forces on nanoparticles~\cite{venkataram2020casimir}.

Finally, as can be seen from the preceding surveys of cloaking, heat 
radiation, light trapping, and optical force limits, the concept of
a scattering cross section is central to a great range of 
electromagnetic problems (others include optical tweezers, laser
heating, etc.); and it is for this reason that it will occupy much of 
our ensuing discussion. For bodies of dimensions $a$ much greater than
$\lambda$, the scattering cross section $\sigma_{sca} \propto a^2$ is
known to scale like the geometric area. 
For electrically small dielectric particles with $a\ll \lambda$, 
the well-known Rayleigh scattering result is $\sigma_{sca}
\propto a^6$. 
Resonant metallic particles in the quasistatic regime provide
larger \emph{relative} optical response $\propto V$, captured in the
aforementioned absorption bounds. As discussed below, recent limit
techniques make it now possible to interpolate between these
asymptotic regimes.
\\ \\
\textbf{General scattering bounds}---A common feature across the 
panoply of electromagnetic limits mentioned so far is the search for 
simplifying assumptions that ``relax'' physical constraints and 
thereby make analysis feasible: working in the geometric optics regime; 
maximizing modal contributions without regards for geometric 
constraints; maximizing material response by application of optical 
theorems based on passivity. 
Over the past few years, a collection of work has formalized the notion 
of physical relaxations through the mathematical language of 
optimization theory, and it is this perspective that will dominate 
the subsequent text. 

Before moving to this topic proper, it is important to recognize that
there are other closely related lines of investigation. 
With regards to antenna design, a great deal of 
progress has been made deriving flexible limits to various aspects of 
antenna performance by formulating the limits as the solutions of 
convex optimizations over possible current distributions of particular 
antenna geometries \cite{gustafsson2013optimal,gustafsson2019maximum,
capek2017minimization,capek2019optimal}. For problems
where an ideal target field distribution $\hat{\vb{E}}(\vb{x})$ is
known, the design may be specified as minimizing the norm squared
deviation $\lVert \vb{E} - \hat{\vb{E}} \rVert^2$ subject to the constraint of
Maxwell’s equations $\mathbb{M}\vb{E}=i\omega\mu_{0}\vb{J}$, with the
Maxwell operator $\mathbb{M}=\curl\curl\, -
k_{0}^2$~\cite{joannopoulosphotonic} and $\vb{J}$ being fixed
sources of the problem. Both the field distribution $\vb{E}(\vb{x})$
and material distribution $\epsilon(\vb{x})$ are then treated as
optimization degrees of freedom, resulting in a non-convex
optimization problem where finding the minimum possible deviation is
computationally difficult~\cite{angeris2021heuristic}. Nevertheless,
much along the lines of what will be done below, the minimum deviation
can be bounded by the global optimum of the convex Lagrangian dual
relaxation~\cite{boyd2004convex}, giving a limit on how closely
$\hat{\vb{E}}$ can be realized in practice; an analogous procedure was
used to obtain bounds on minimum achievable mode volumes of dielectric
resonators, given constraints on device size and
material~\cite{zhao2020minimum}. More broadly, the method can be
extended to any separable functions of the field at different
positions $f(\vb{E}) = \sum_{\vb{x}}f_{\vb{x}}(\vb{E}(\vb{x}))$, which
are of great relevance to any design problem concerning the
actualizing of some specific field transformations. For other types
of objectives, especially if there is no rigorous way to assert that
some particular field solution is optimal, it may be difficult to
actually evaluate the form of the dual function and thereby obtain
limits~\cite{angeris2021heuristic}.
\section*{Technical description}
\noindent
\textbf{Scattering preliminaries}---As noted above, the core idea 
motivating the study of physical limits is to extract attributes that 
apply to many realizable instances (devices with different material 
parameters, structural parameters, etc.) by means of some relaxation: 
when the space of possible designs (or solutions) is too complex to be 
characterized directly, as is almost always the case, the only tractable 
approach to understanding the degree to which universal properties may 
be controlled is to equate large groups of designs by purposefully 
filtering out certain details. 
To this end, the perspective offered by scattering theory 
is quite helpful. 
First, by working from the basic definitions of scattering theory, 
the relationship between the structure of the potential and the 
polarization density generated by a particular excitation becomes 
more apparent. 
Second, as scattering descriptions innately lead to integral 
formulations, the constraints of any scattering theory are naturally 
organized into a spatial hierarchy that meshes well with both 
physical intuition and optimization. 
These two aspects, taken together, establish the crucial link 
between the features that may be imparted to waves through material 
structuring, and the standard forms and techniques 
of optimization theory.
Throughout the following text, the constitutive relations
$\textbf{D}\left(\textbf{x},\omega\right) =
\epsilon_{0}\mathbb{V}\left(\textbf{x},\omega\right)\cdot
\textbf{E}\left(\textbf{x},\omega\right)$ and $\textbf{B}\left(
\textbf{x},\omega\right) = \mu_{0}
\textbf{H}\left(\textbf{x},\omega\right)$ will be assumed for 
simplicity. 
However, much of the subsequent development can be carried out 
in greater generality, e.g. magnetic media, non-reciprocal media, etc.,
c.f. Ref.~\cite{kuang2020computational}.

\begin{figure}[t!]
\centering
\includegraphics[width=1\columnwidth]{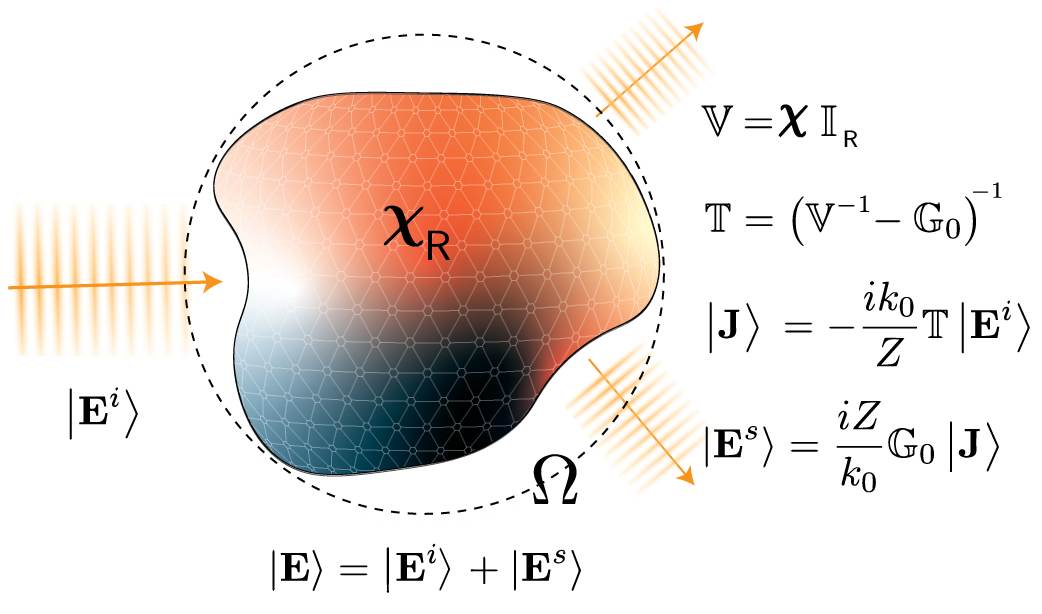}
\vspace{-10 pt}
\caption{\textbf{Schematic of scattering theory.} The basis of 
scattering theory broadly rests on the existence, for each 
particular object (scattering potential), of an exact 
relation (the $\mathbb{T}$-operator) between incident 
(initial or free) electromagnetic fields and total 
(generated or net) polarization currents. 
Once this relation is known, the total electromagnetic field 
(the solution of Maxwell's equations) is given by 
applying the Green's function of the free (empty) domain 
$\mathbb{G}_{0}$ to the total current associated with 
the incident field.}
\end{figure}

At the finest level of detail that the design of a photonic device may 
be described (``true physics''), Maxwell's wave equation associates
each unique inhomogeneous dielectric profile $\mathbb{V}\left(
\textbf{x},\omega\right)$ within a given domain $\Omega$ 
(assumed to enforce outgoing scattered fields~\cite{berenger2007huygens,
lindell2019boundary}), with a unique differential equation,
\begin{equation}
 \nabla\times\nabla\times
 \textbf{E}\left(\textbf{x},\omega\right)
 - k_{0}^{2}~\mathbb{V}\left(\textbf{x},\omega\right)\cdot
 \textbf{E}\left(\textbf{x},\omega\right)
 = i\omega\mu_{0}~\textbf{J}^{f}\left(\textbf{x},\omega\right), 
 \label{MaxwellWave}
\end{equation}
where $k_{0} = \omega / c = 2\pi/\lambda$ and $\lambda$ is the free
space wavelength. 
Unless $\mathbb{V}\left(\textbf{x},\omega\right)$ exhibits special
symmetries, equations like \eqref{MaxwellWave} do not typically have
complete closed form solutions, meaning that it is usually not
possible to fully analyze how changes to the potential $\mathbb{V}$
alter the total field $\textbf{E}$ beyond local expansions
(derivatives)~\cite{molesky2020t}. Nevertheless, this does not mean
that the relation between $\mathbb{V}$ and $\textbf{E}$ is completely
opaque. In particular, suppose that $\textbf{J}^{f}\left(\textbf{x},
\omega\right)= \textbf{0}$, and that the boundary conditions on $\Omega$ 
are set to describe an incident (incoming) electromagnetic field with 
electric component $\textbf{E}^{i}\left(\textbf{x},\omega\right)$. 
Introducing $\mathbb{I}_{_\Omega}\left(\textbf{x},\omega\right)$ as 
the identity operator, $\mathbb{I}_{_\Omega}\cdot\textbf{A} = 
\textbf{A}$, Eq.~\eqref{MaxwellWave} may be rewritten as 
\begin{align}
 &\nabla\times\nabla\times\textbf{E}\left(\textbf{x},\omega\right)
 - k_{0}^{2}~\mathbb{I}_{_\Omega}\left(\textbf{x},\omega\right)\cdot
 \textbf{E}\left(\textbf{x},\omega\right) =
 \nonumber \\
 &k_{0}^{2}~\left(\mathbb{V}\left(\textbf{x},\omega\right)
 -\mathbb{I}_{_\Omega}\left(\textbf{x},\omega\right)\right)\cdot
 \textbf{E}\left(\textbf{x},\omega\right).
 \label{scat1}
\end{align}
The scheme of Eq.~\eqref{scat1} is no different than that of 
Eq.~\eqref{MaxwellWave} as applied to vacuum and the polarization 
``source'' $\left(-i k_{0}/Z\right)$ $\left(\mathbb{V}\left(\textbf{x},
\omega\right)-\mathbb{I}\left(\textbf{x},\omega\right)\right)\cdot
\textbf{E}\left(\textbf{x},\omega\right)$, and as such, 
Eq.~\eqref{scat1} implies an implicit integral relation on 
$\textbf{E}$ (a volume integral 
formulation~\cite{sun2009novel,samokhin2013integral}) that, while 
offering a distinct conceptual perspective~\cite{costabel2012essential} 
and possible computational advantages~\cite{polimeridis2014computation,
polimeridis2015fluctuating,liu2018potential}, is functionally 
equivalent to Maxwell's equations. 
Setting $\mathbb{R}\left(\textbf{x},\omega\right) =
\mathbb{V}\left(\textbf{x},\omega\right) - \mathbb{I}\left(\textbf{x},
\omega\right)$ and decomposing $\textbf{E}\left(\textbf{x},\omega
\right)$ as $\textbf{E}\left(\textbf{x},\omega\right) = \textbf{E}^{i}
\left(\textbf{x},\omega\right) + \textbf{E}^{s}\left(\textbf{x},\omega
\right)$ (with the $i$ superscript standing for ``incident'' and the 
$s$ superscript for ``scattered''), regardless of what the shape 
specified by $\mathbb{R}\left(\textbf{x},\omega\right)$ actually is, 
$\textbf{E}\left(\textbf{x},\omega\right)$ must obey the 
self-referential (Lippmann-Schwinger~\cite{lippmann1950variational}
or Liouville-Neumann series relation~\cite{lanczos1950iteration,
tsang2004scattering})~\footnote{It is equally possible to derive results 
completely analogous to Eqs.~\eqref{tOne}--\eqref{tFour} using 
scattered electromagnetic fields~\cite{angeris2021heuristic}, 
as opposed to the polarization current density perspective used here.}
\begin{align}
 \textbf{E}^{s}\left(\textbf{x},\omega\right)
 &= 
 \int\limits_{_\Omega}\!\!d\textbf{x}'~
 \mathbb{G}_{0}\left(\textbf{x},\textbf{x}',\omega\right)\cdot 
 \mathbb{R}\left(\textbf{x}',\omega\right)\cdot
 \textbf{E}\left(\textbf{x}',\omega\right)
 \label{scat2}\\
 &=
 \int\limits_{_\Omega}\!\!
 \mathbb{G}_{0}\cdot 
 \mathbb{R}\cdot\textbf{E}^{i} 
 +
 \iint\limits_{_\Omega}\!\!
 \mathbb{G}_{0}\cdot 
 \mathbb{R}\cdot\mathbb{G}_{0}\cdot 
 \mathbb{R}\cdot\textbf{E}^{i} + \dots, 
 \nonumber
\end{align}
where, taking $\textbf{r}$ to be $k_{0}\left(\textbf{x}- \textbf{x}'
\right)$, $r = \lVert\textbf{r}\rVert$, $\hat{\textbf{r}}\otimes
\hat{\textbf{r}}$ to be the vector outer product of $\hat{\textbf{r}} = 
\textbf{r}/\lVert\textbf{r}\rVert$ with itself, and
$\overline{\textbf{id}}$ the $3\times 3$ vector identity matrix, 
\begin{align}
 &\mathbb{G}_{0}\left(\textbf{x}-\textbf{x}',\omega\right)
 = 
 \nonumber \\
 &\frac{k_{0}^{3}}{4\pi}\frac{e^{i r}}{r}
 \Bigg[\left(1 + \frac{i r-1}{r^{2}}\right)
 \overline{\textbf{id}} ~- 
 \left(1 + 3\frac{ir - 1}{r^{2}}\right)
 \hat{\textbf{r}}\otimes\hat{\textbf{r}}\Bigg],
 \label{greenForm}
\end{align}
is the vacuum Green's function for the left hand side of
Eq.~\eqref{scat1}~\footnote{The 
Green's function (fundamental solution) of a linear differential 
equation is the solution of the differential equations for the Dirac 
distribution, i.e. $\left(1/k_{0}^{2}\right)\nabla\times\nabla\times
\mathbb{G}_{0}\left(\textbf{x},\textbf{x}',\omega\right)-\mathbb{G}_{0}
\left(\textbf{x},\textbf{x}',\omega\right) = \bm{\delta}\left(\textbf{x}-
\textbf{x}'\right)$ in our notation. 
The solution of the differential equation for any inhomogeneous source
term is given by convolution with the Green's 
function~\cite{novotny2012principles}.}. 
(Note that an the additional factor of $k_{0}^{2}$ 
is included in Eq.~\eqref{greenForm} compared to its usual 
definition~\cite{novotny2012principles}. 
This is done so that every length that appears in the use of the 
Green's function inside volume integrals is defined relative to the 
wavelength.) 
The mathematical form of Eq.~\eqref{scat2} abstractly shows that
$\textbf{J}\left(\textbf{x},\omega\right)
= -\frac{i k_{0}}{Z}\mathbb{R}\left(\textbf{x},
\omega\right)\cdot\textbf{E}\left(\textbf{x},\omega\right)$ must be 
regarded as the \emph{total} polarization current density generated in
response to $\textbf{E}^{i}\left(\textbf{x},\omega\right)$, and that 
Eq.~\eqref{scat1} may be equivalently stated (in Fredholm integral 
form~\cite{samokhin2013integral,kanwal2013linear}) as
\begin{align}
 &\textbf{J}\left(\textbf{x},\omega\right) - 
 \int\limits_{_\Omega}
 \!\!d\textbf{x}'~
 \mathbb{R}\left(\textbf{x},\omega\right)
 \cdot\mathbb{G}_{0}\left(\textbf{x},\textbf{x}',
 \omega\right)\cdot \textbf{J}\left(\textbf{x}',\omega\right) =
 \nonumber  \\
 &-\frac{i k_{0}}{Z}~\mathbb{R}\left(\textbf{x},\omega\right)
 \cdot\textbf{E}^{i}\left(\textbf{x}',\omega\right)
 = \textbf{J}^{i}\left(\textbf{x},\omega\right),
 \label{tOne}
\end{align}
where $\vec{J}^{i}$ is the ``initial'' polarization current density
setup in response to the initial electric field, see Fig.~3.

Eqs.~\eqref{scat2} and \eqref{tOne} rest at the foundation of 
scattering theory and definition of the 
$\mathbb{T}$-operator~\cite{kruger2012trace,tsang2004scattering} 
($\mathbb{S}$-matrix~\cite{dyson1949s,gell1953formal,van1953s}). 
Directly, via Eq.~\eqref{scat2}, every incident electric field
is related to a specific polarization current density (the
polarization that it generates) by
\begin{align}
 &\mathbb{I}_{_{\mathsf{R}}}
 \textbf{E}^{i}\left(\textbf{x},\omega\right) =
 \label{tOpt1} \\
 &\!\int\limits_{\mathsf{R}}\!d\textbf{x}'~
 \left(\bm{\delta}\left(\textbf{x}-\textbf{x}'\right)
 - \mathbb{G}_{0}\left(\textbf{x},\textbf{x}', \omega\right)
 \cdot \mathbb{R}\left(\textbf{x}',\omega\right)\right)
 \cdot\textbf{E}\left(\textbf{x}',\omega\right) =
 \nonumber \\
 &\!\frac{iZ}{k_{0}}\int\limits_{\mathsf{R}}\!d\textbf{x}'~
 \left(\mathbb{R}^{-1}\left(\textbf{x}',\omega\right)
 - \mathbb{G}_{0}\left(\textbf{x},\textbf{x}',
 \omega\right)\right) \cdot\textbf{J}\left(\textbf{x}',\omega\right),
 \nonumber
 \end{align} 
where $\mathsf{R}$ is the subdomain of $\Omega$ where
$\mathbb{R}\left(\textbf{x},\omega\right) \neq \textbf{0}$, the
``material'' extent of the scattering potential, and
$\mathbb{R}^{-1}\left(\textbf{x},\omega\right)$ is the pseudo inverse
of $\mathbb{R}$, the inverse over the subdomain where $\mathbb{R}$ is
nonzero. 
Because $\mathbb{R}$ and $\mathbb{G}$ describe causal,
passive, linear system responses, the total linear operator relating
$\textbf{J}$ to $\textbf{E}^{i}$ within $\mathsf{R}$ must be
``invertible''~\cite{landau2013statistical,rudin1991functional}.
Accordingly, Eq.~\eqref{scat2} delineates the existence of the inverse
relation
\begin{align}
  &\textbf{J}\left(\textbf{x},\omega\right) = 
  -\frac{ik_{0}}{Z}\int_{_\mathsf{R}}d\textbf{x}'~\mathbb{T}\left(
  \textbf{x},\textbf{x}',\omega\right)\cdot\textbf{E}^{i}
  \left(\textbf{x}',\omega\right) \Rightarrow
  \label{tOpt2}\\
  &\mathbb{I}_{_{\mathsf{R}}}=
  \int\limits_{\mathsf{R}}d\textbf{x}'~
  \left(\mathbb{R}^{-1}\left(\textbf{x}',\omega\right)
  -\mathbb{G}_{0}\left(\textbf{x},\textbf{x}',
  \omega\right)\right) \cdot
  \mathbb{T}
  \left(\textbf{x}',\textbf{x}'',\omega\right),
  \nonumber 
\end{align}
defining for each unique scattering potential $\mathbb{V}$ a
unique linear response function $\mathbb{T}$ that relates any incident
field $\textbf{E}^{i}$ to the polarization current density $\textbf{J}$ 
that self-consistently solves Maxwell's equations through
Eq.~\eqref{scat2}. 
The operator relation governing the $\mathbb{T}$-operator,
the Green's function, and $\mathbb{V}$ given in Eq.~\eqref{tOpt2}, like
Eqs.~\eqref{scat2} and \eqref{tOpt1}, is fully equivalent to Maxwell's
equations and serve as an advantageous starting point for deriving
conserved quantities.

From Eq.~\eqref{tOpt1}, relatively little must be done to reframe the
determination of an optimal scattering object $\mathbb{V}$ in terms of
the polarization current density. 
First, by integrating against the characteristic function of any
other \emph{known} subdomain $\mathsf{P}$ of $\Omega$, together with a
local ``polarization'' projection matrix
$\mathbb{P}\left(\textbf{x},\omega\right)$---a linear response that
does not mix distinct spatial points---the integration domains
appearing in Eq.~\eqref{tOne} are shifted from $\mathsf{P}$ to
$\mathsf{R}\cap\mathsf{P}$, standing for the common spatial volume of
$\mathsf{R}$ and $\mathsf{P}$. 
Next, by applying this result against
the conjugate of $\textbf{J}\left(\textbf{x},\omega\right)$ itself,
for \emph{any} $\mathbb{P}\left(\textbf{x},\omega\right)$
Eq.~\eqref{tOpt1} is transformed into
\begin{multline}
 \int\limits_{_{\mathsf{P}}}
 \!\!d\textbf{x}~~
 \textbf{J}^{*}\left(\textbf{x},\omega\right) 
 \cdot
 \mathbb{P}\left(\textbf{x},\omega\right)
 \cdot \\ \left(\underline{\mathbb{R}}^{-1}\left(\textbf{x},
 \omega\right)
 \cdot \textbf{J}\left(\textbf{x},\omega\right) 
 - \int\limits_{_\Omega}\mathbb{G}_{0}
 \left(\textbf{x}-\textbf{x}',
 \omega\right) \cdot 
 \textbf{J}\left(\textbf{x}',\omega\right)\right) =
 \\  
 -\frac{i k_{0}}{Z}\int\limits_{_{\mathsf{P}}}
 \!\!d\textbf{x}~
 \textbf{J}^{*}\left(\textbf{x},\omega\right) 
 \cdot
 \mathbb{P}\left(\textbf{x},\omega\right)
 \cdot\textbf{E}^{i}\left(\textbf{x}\right); 
 \label{tFour}
\end{multline}
where, crucially, the dependence of the domains of integration on the 
spatial structure of $\mathbb{R}\left(\textbf{x},\omega\right)$ has been 
removed. 
Based on the fact that $\mathbb{R}\left(\textbf{x},\omega\right) = 
\textbf{0}$ implies that $\textbf{J}\left(\textbf{x},\omega\right) = 
\textbf{0}$, the content of Eq.~\eqref{tFour} is unchanged 
for \emph{any} choice of response function $\underline{\mathbb{R}}^{-1}$ 
on $\Omega$ so long as $\underline{\mathbb{R}}^{-1}\left(\textbf{x},
\omega\right) = \mathbb{R}^{-1}\left(\textbf{x},\omega\right)$ when 
$\textbf{x}\in\mathsf{R}$. 
If $\mathbb{V}\left(\textbf{x},\omega\right)$ may only take on a 
single $3\times 3$ matrix form distinct from $\textbf{0}$, 
as is usually true in photonics when designing a device composed 
of a single material~\footnote{The same conclusion, with minor 
modifications, also holds if the design domain $\Omega$ is split 
into a collection of subdomains, and in each subdomain any possible 
structuring must be carried out in a single known material.}, 
then Eq.~\eqref{tFour}, with $\bm{\chi}\left(\omega\right)$ 
denoting the electric susceptibility matrix of the material, 
may be further simplified by setting 
$\underline{\mathbb{R}}^{-1}\left(\textbf{x},\omega\right) 
=\bm{\chi}^{-1}\left(\textbf{x},\omega\right)$, see 
Eq.~\eqref{braKetConstraints} below. 
The only quantity in Eq.~\eqref{tFour} that is not typically known from 
the outset of a design problem is $\textbf{J}\left(\textbf{x},\omega\right)$. 
At the same time, because the physics of Maxwell's equation 
is incorporated through $\mathbb{G}_{0}$, the true 
behaviour of the dielectric scattering potential is incorporated through 
$\bm{\chi}^{-1}$, and the self-consistency of viewing $\textbf{J}
\left(\textbf{x},\omega\right)$ as the electric polarization current 
density resulting from the electromagnetic field given by 
$\textbf{E}^{i}\left(\textbf{x},\omega\right)$ is incorporated by 
association with Eq.~\eqref{scat2}, \emph{any} vector field that 
respects Eq.~\eqref{tFour} for all possible choices of $\mathbb{P}\left(
\textbf{x},\omega\right)$ actually defines an effective medium 
scattering structure~\cite{molesky2021comm} (a mix between the
material properties asserted by
$\mathbb{V}\left(\textbf{x},\omega\right)$ and the background). 
\\ \\
\textbf{Optimization bounds}---The observation that every constraint 
of the form given by Eq.~\eqref{tFour} applies to \emph{any} 
structure of a given material within $\Omega$ implies that a great 
number of common photonic objectives can be 
stated as quadratically constrained quadratic programs 
(QCQPs)~\cite{angeris2021heuristic}, and, in turn,
bounded using standard relaxation techniques from optimization
theory~\cite{boyd2004convex}. 
The connection between Eq.~\eqref{tFour} and QCQPs is most easily seen 
by switching to a more compact notation. 
Making use of the fact that integration may be viewed as an inner product 
for fields or functions~\cite{rudin2006real}~\footnote{Technically, 
the integral is an inner product for almost everywhere equal equivalences 
classes of functions and fields. However, because we are primarily 
concerned with finite numerical representations, here the distinction is not 
important.}, with $\left<\textbf{A}\big|\textbf{B}\right> = \int
d\vec{x}~\textbf{A}^*(\vec{x})\cdot \textbf{B}(\vec{x})$ denoting the
standard complex-conjugate inner product, Eq.~\eqref{tFour} may be
written in bra-ket form as
\begin{equation}
 \left<\textbf{T}\right|\mathbb{P}
 \left(\bm{\chi}^{-1}
 - \mathbb{G}_{0}\right) 
 \left|\textbf{T}\right> = 
 \left<\textbf{T}\right|\mathbb{P}
 \left|\textbf{E}^{i}\right>,
 \label{braKetRel}
\end{equation}
with $\left|\textbf{T}\right> = 
\frac{i Z}{k_{0}}\left|\textbf{J}\right> = \mathbb{T}\left|
\textbf{E}^{i}\right>$, leading to the following equivalent adjoint 
quadratic constraint
equations
\begin{align}
 &\Im\left(\left<\textbf{E}^{i}\right|\mathbb{P}\left|
 \textbf{T}\right>\right) 
 - 
 \left<\textbf{T}\right|
 \left[\left(\bm{\chi}^{-1\dagger}
 - \mathbb{G}_{0}^{\dagger}
 \right)\mathbb{P}\right]^{\mathsf{A}}\left|\textbf{T}\right> = 0,
 \nonumber \\
 &\Re\left(\left<\textbf{E}^{i}\right|\mathbb{P}\left|
 \textbf{T}\right>\right) 
 - 
 \left<\textbf{T}\right|
 \left[\left(\bm{\chi}^{-1\dagger}
 - \mathbb{G}_{0}^{\dagger}
 \right)\mathbb{P}\right]^{\mathsf{S}}\left|\textbf{T}\right> = 0,
 \label{braKetConstraints}
\end{align}
In these expressions, and the proceeding text, $\mathsf{S}$ 
superscripts will be used to mark the Hermitian 
(symmetric) part of the contained linear response function 
$\textbf{M}^{\mathsf{S}} = 
\left(\textbf{M} + \textbf{M}^{\dagger}\right)/2$, and
$\mathsf{A}$ superscripts will be used to denote the anti-symmetric 
part, $\textbf{M}^{\mathsf{A}} = 
\left(\textbf{M} - \textbf{M}^{\dagger}\right)/2i$, 
so that, like a complex number, $\textbf{M} = \textbf{M}^{\mathsf{S}} +
i~\textbf{M}^{\mathsf{A}}$~\footnote{Given our freedom in choosing
$\mathbb{P},$ there is no difference between $\mathbb{P}$ and
$\mathbb{P}^{\dagger}$.}. 

When $\mathbb{P}$ is set to the domain identity $\mathbb{I}_{_\Omega}$, 
the first relation of Eq.~\eqref{braKetConstraints} is a statement of 
the conservation of \emph{real} power within the 
domain~\cite{kuang2020maximal}: 
the power drawn by the polarization current from the field, 
the inner product 
$\Im\left(\left<\textbf{E}^{i} \big| \textbf{T}\right>
\right)$, must equal the sum of the power lost by the polarization 
current to material extinction~\cite{miller2014fundamental},
\begin{equation}
\left<\textbf{T}\right| \left(\bm{\chi}^{-1\dagger}\right)^{\mathsf{A}} 
\left|\textbf{T}\right> = \!\int\limits_{\Omega}\!\!d\textbf{x}
~\textbf{T}^{*}
 \left(\textbf{x}\right)\cdot\frac{\Im\left[\mathbb{V}\left(
 \textbf{x},\omega\right)\right]~}{\lVert\mathbb{V}\left(\textbf{x},
 \omega\right)\rVert^{2}}\cdot\textbf{T}\left(\textbf{x}\right)
 \label{matLoss}
\end{equation}
and to outgoing radiation~\cite{molesky2019bounds}
\begin{equation}
\left<\textbf{T}\right| \mathbb{G}_{0}^{\mathsf{A}} 
\left|\textbf{T}\right> = 
\!\iint\limits_{\Omega}\!\!d\textbf{x}'d\textbf{x}~\textbf{T}^{*} 
\left(\textbf{x}'\right)\cdot\Im\left[\mathbb{G}_{0}
\left( \textbf{x}',\textbf{x},\omega\right)\right]\cdot 
\textbf{T}\left(\textbf{x}\right). 
\label{radLoss}
\end{equation}
The second relation contained in Eq.~\eqref{braKetConstraints}, as
illustrated further below, makes the related statement
that \emph{reactive} power must also be conserved, see
Refs.~\cite{jackson1999classical, gustafsson2020upper,molesky2020t}.
Together, these equalities impose minimum requirements on the
characteristics of $\mathbb{V}$ and spatial extent of the domain
$\Omega$---through $\mathbb{G}_{0}$ which is limited to
$\Omega$---needed to achieve resonant response. That is, depending on
the material the device will be made of and the spatial volume that it
may possibly occupy, there may be strong limits on the degree to which
the amplitude and phase of $\textbf{J}$ relative to $\textbf{E}^{i}$
may be tuned.

Recalling from Eq.~\eqref{scat2} that the field and induced
polarization currents are linearly related by $\left|\textbf{E}\right>
= \left|\textbf{E}^{i}\right>
+ \frac{iZ}{k_{0}}\mathbb{G}_{0}\left|\textbf{J}\right>$, it follows that any
physical process described by quadratic field terms---including the
fundamental time-averaged power-transfer quantities of absorption
$k_{0} \left<\textbf{E}\right|
\bm{\chi}^{\mathsf{A}}\left|\textbf{E}\right>/2Z$, extraction 
$\Re\left(\left<\textbf{E}^{i}\big|\textbf{J}\right>\right)/2$ 
and scattering $Z\left<\textbf{J}\right|
\mathbb{G}_{0}^{\mathsf{A}}\left|\textbf{J}\right>/2k_{0}$, which 
rest as the basic figures of merit for the design of 
antennas~\cite{vercruysse2014directional,
shahpari2018fundamental,capek2019optimal}, light trapping
devices~\cite{yablonovitch1982statistical,
siegel1993refractive,yu2012thermodynamic,callahan2012solar,
mokkapati2012nanophotonic,miroshnichenko2018ultimate}, and
optoelectronic coupling~\cite{niv2012near,miller2013photon,
xu2015generalized,liu2016fundamental}---can be considered as 
a quadratic objective. 
In this language, taking $\mathtt{f}_{0}\left(\left|\textbf{T}\right>
\right)$ to denote some quadratic function of the 
polarization, and $\text{K}$ a complete set of
constraints~\footnote{Practically, the size of $\text{K}$ in numeric
simulation is set by the size of the fields.} the goal of maximizing
(resp. minimizing) any such objective through material structuring may
be formulated as
\begin{align}
 &\max_{\left|\textbf{T}\right>}~\left(\text{resp.}
 \min_{\left|\textbf{T}\right>}\right)~
 \mathtt{f}_{0}\left(\left|\textbf{T}\right>\right)
 \nonumber \\
 &\text{such that}~\forall k\in \text{K}
 \nonumber \\
 &\Re\left(\left<\textbf{E}\right|\mathbb{P}_{k}
 \left|\textbf{T}\right>\right) 
 - 
 \left<\textbf{T}\right|
 \left[\mathbb{P}_{k}\left(\bm{\chi}^{-1\dagger}
 -\mathbb{G}_{0}^{\dagger}\right)\right]^{\mathsf{S}}\left|
 \textbf{T}\right> = 0,
 \nonumber \\
 &\Im\left(\left<\textbf{E}\right|\mathbb{P}_{k}
 \left|\textbf{T}\right>\right) - 
 \left<\textbf{T}\right|
 \left[\mathbb{P}_{k}\left(\bm{\chi}^{-1\dagger}
 -\mathbb{G}_{0}^{\dagger}\right)\right]^{\mathsf{A}}
 \left|\textbf{T}\right> = 0,
 \label{qcqp}
\end{align}
which is the form of a quadratically constrained quadratic program
(QCQP). 
Because the enforcement of fewer constraints always leads to
maxima of greater or equal value (resp. minima of equal or smaller
value) in any optimization, the imposition of any collection of
constraints that can be formed from $\text{K}$ may be used to
construct a relaxed QCQP that contains the feasible set of
Eq.~\eqref{qcqp}---an optimization with maxima (resp. minima) at least
as large (resp. small) as Eq.~\eqref{qcqp}~\footnote{A
feasible field is a field that respects every imposed constraint. The
feasible set of an optimization statement is the collection of all
feasible fields.}. Any bound on such a
relaxed program is necessarily a bound on Eq.~\eqref{qcqp}, and so, by
applying any additional relaxation such as Lagrange duality or
semi-definite programming~\cite{angeris2021heuristic}, it is possible
to obtain limits on the physically realizable values of
$\mathtt{f}_{0}\left(\left|\textbf{T}\right>\right)$ that universally
apply to any possible material structure within $\Omega$,
c.f. Refs.~\cite{gustafsson2020upper,molesky2020t,kuang2020maximal,
molesky2020hierarchical,kuang2020computational,schab2020trade,
capek2021fundamental,molesky2021comm,angeris2021heuristic}. 
As highlighted by the expositive examples below, the extent to which
these limits incorporate various physical phenomena may be tuned by
selecting, either by intuition or algorithm~\cite{kuang2020maximal},
the collection of constraints ($\mathbb{P}_{k}$ projections) that are
concurrently imposed, and, in contrast to many traditional approaches 
to limits, where individual components of an expression are bounded 
and then subsequently summed or composed to form a global bound, 
the optimization framework of Eq.~\eqref{qcqp} properly describes 
interactions between constraints. 

Although no further refinements of Eq.~\eqref{qcqp} will be examined
hereafter, it should be noted that this basic optimization bounds 
approach can, and in certain cases should, be extended in at least two
meaningful ways. First, by moving to complex frequencies as described
in Ref.~\cite{liang2013formulation} and
Ref.~\cite{shim2019fundamental}, it is possible to adapt
Eq.~\eqref{qcqp} to treat both broadband and temporal problems.
Detailed accounts of these modifications can be found in
Refs.~\cite{kuang2020computational,zhang2021conservationlawbased}.
Second, in situations involving multiple incident fields or scattering
objectives, including applications to multi-functional
devices---design objectives like optical 
multiplexing~\cite{li2017recent,yang2020density,
feng2020wavelength}, meta-optic imaging
components~\cite{staude2019all,lin2020end,phan2019high,
christiansen2020fullwave} and optical
computing~\cite{estakhri2019inverse,li2019intelligent,
rajabalipanah2020space}---it is necessary to broaden the scope of the
quadratic equalities included in Eq.~\eqref{qcqp} to properly account
for the additional challenges presented by the need to engineer
multiple field transformations within a common structure. 
A full account of these alterations is given in
Ref.~\cite{molesky2021comm}.
\section*{Representative scattering limits}
In order to build intuition and provide context, the ensuing section
reviews three increasingly complex tutorial applications of
Eq.~\eqref{qcqp} to set fundamental upper bounds on optical response. 
Beginning with the conservation of \emph{real power}, focusing on the 
equivalent problems of maximizing thermal emission or net
absorption, the modal characteristics of $\mathbb{G}_{0}^{\mathsf{A}}$
in relation to Eq.~\eqref{braKetConstraints} are shown to reproduce
familiar asymptotic results from quasi-statics and ray optics. 
However, because optimization limits do not rely on the validity of 
such approximate forms, calculated bounds are also seen to be 
meaningfully applicable to intermediate and hitherto inaccessible 
wavelength-scale regimes. 
Next, by further imposing that \emph{reactive power} be conserved,
simultaneously enforcing the two constraints in
Eq.~\eqref{braKetConstraints} through $\mathbb{P} =
\mathbb{I}_{_{\Omega}}$, limits on achievable scattering cross 
sections are found to anticipate conditions on the size of the design
domain under which resonant response is possible for a given material
choice. Finally---as exemplified through calculations of bounds on 
scattering cross sections, radiative emission from a dipolar source 
in the near field of body, and power splitting---the set of integral 
relations contained in the relaxations of Eq.~\eqref{qcqp} has the 
effect of defining the degree to which the physics of scattering theory 
is enforced, and correspondingly, the number and types of integral 
constraints imposed in calculating optimization bounds function as 
complements to the different number and types of optimization degrees 
of freedom that may be used in structural optimization. 
For proper comparison against realizable devices, constraints must be 
selected in a manner that ``resolves'' the wave physics of the problem, 
e.g., accurate limits on phenomena dominated by rapidly decaying (evanescent) 
fields require a greater number of local constraints. 
In almost all of these representative applications, objective
values obtained through structural ``topology'' (or ``density'')
optimization are found to come within an order of magnitude of their
determined limit values~\cite{christiansen2021inverse}.
\\ \\
\textbf{Real power conservation}---As a first application of
Eq.~\eqref{qcqp}, we review how the conservation of real power sets an
upper bound on thermal radiation and, by reciprocity, angle-integrated
absorption. At a microscopic level, thermal emission results from
stochastically fluctuating electrical currents in
matter~\cite{kravtsov1975statistical,rytov1988principles2}, with the
precise relationship between temperature, energy dissipation, and
field fluctuations in an object in equilibrium determined by the
fluctuation-dissipation
theorem~\cite{kubo1966fluctuation}. 
The basis for such a relation may be intuitively understood from Brownian
motion~\cite{morters2010brownian}. 
A particle traveling in a fluid
experiences a dissipation of its net velocity due to collisions with
the constituent particles of the surrounding fluid. 
Complementarily, these collisions impart momentum, causing fluctuations 
in the position of the particle about its average location, 
$\langle x^{2} \rangle = 2Dt$, where $t$ is the elapsed time and 
$D$ is the diffusion coefficient, which by the 
Stokes-Einstein relation $D = k_{B}T/\gamma$ is inversely 
proportional to the the drag (dissipation) 
coefficient $\gamma$~\cite{marconi2008fluctuation}. 
An analogous relation is seen in the Nyquist formula for
Johnson noise, $\langle V^{2} \rangle = 4Rk_{B}T d\nu$, where $V$ is
the voltage between the terminals of an open circuit (e.g. a
conductive wire), $R$ the electrical resistance, and $d\nu$ a
frequency interval~\cite{landauer1989johnson}. 
The fluctuation-dissipation theorem generalizes and formalizes 
these observations. For the electromagnetic settings 
considered here~\cite{novotny2012principles}, 
\begin{align}
 &\underbrace{\left\{ J_{i}(\mathbf{x}, \omega)
 J_{j}^{*}(\mathbf{x}', \omega ') 
 \right\}_{T}}_{\text{fluctuation}}=
 \label{fluctuationDissipation}
 \\ &\frac{\omega\epsilon_{0}}{2\pi} 
 \coth\left( \frac{\hslash
 \omega}{2k_{B}T} \right)
 \underbrace{\bm{\chi}_{ij}^{\mathsf{A}}(\mathbf{x},\omega)
 }_{\text{dissipation}}
 ~\delta_{ij}~\delta(\mathbf{x} - \mathbf{x}')~
 \delta(\omega - \omega'),
 \nonumber
\end{align}
where $\left\{\dots\right\}$ denotes a thermal ensemble average:
Fluctuations in the current density are point correlated 
and proportional to the dissipative part of the electric 
susceptibility (the optical conductivity).

\begin{figure*}[t]
 \includegraphics[width=1.0\textwidth]{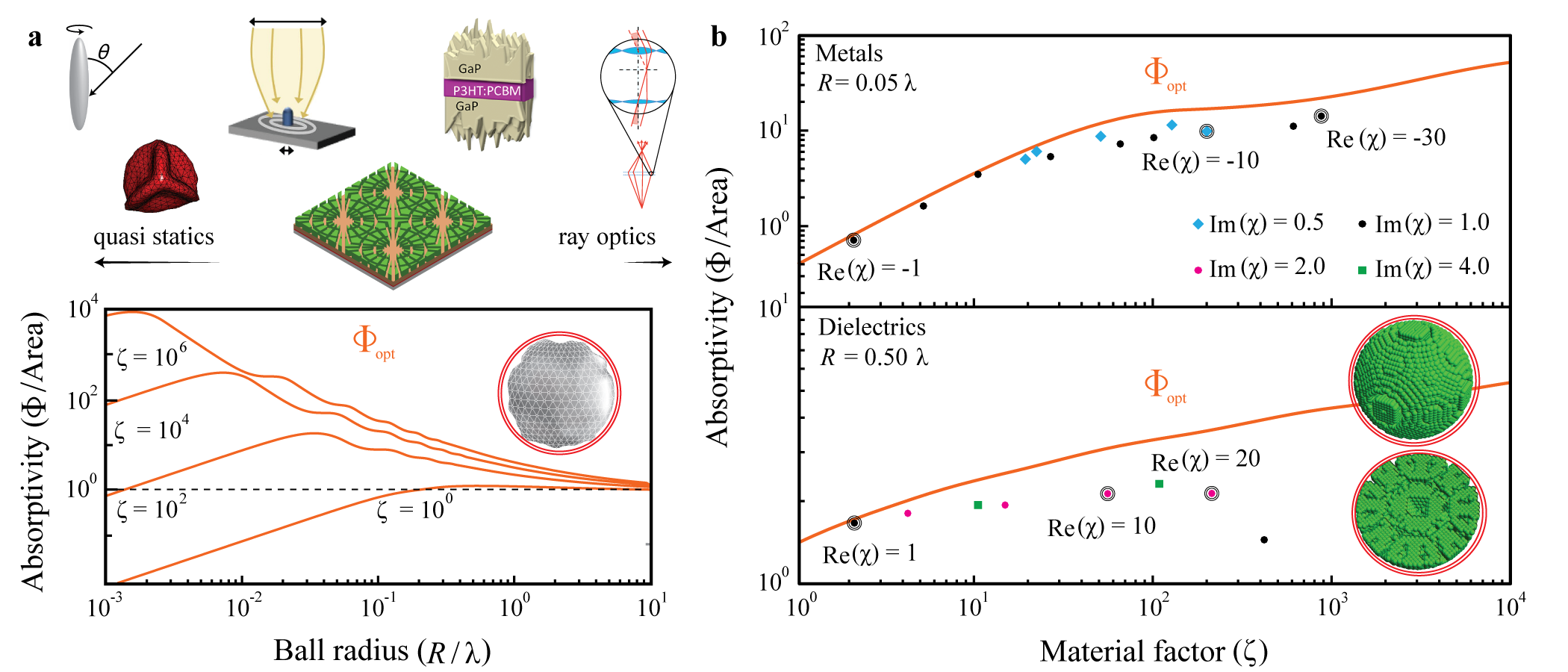} 
 \caption{\textbf{Bounds on angle-integrated absorption based on the 
 conservation of real power.} Panel \textbf{a} illustrates the results of
 Eq.~\eqref{realPowerProb} as a function of the size (radius
 $R/\lambda$) of a spherically bounded design volume for several
 values of the material factor $\zeta
 = \lVert\chi\rVert^{2}/\Im\left(\chi\right)$. Notably, in
 transitioning between very small design volumes, $R\ll \lambda$, and
 very large design volumes, $R\gg\lambda$, Eq.~\eqref{realPowerProb}
 is seen to smoothly blend the familiar $\propto V$ and $\propto A$
 limits of quasi-static and ray-optic (blackbody) approximations.
 Above panel \textbf{a}, a collection of representative use cases of
 radiative absorption are included to give a sense of the range of
 length scales covered below. Working from left to right, these
 images are taken from Ref.~\cite{gustafsson2007physical},
 Ref.~\cite{miller2014fundamental}, Ref.~\cite{xu2015generalized},
 Ref.~\cite{callahan2012solar}, and
 Ref.~\cite{hamilton2009metamaterials}. Panel \textbf{b} depicts a
 comparison of $\Phi_{opt}$ against performance values obtained by
 computational design methods for two set domain sizes, $R/\lambda =
 1/20$ for metals ($\Re\left(\chi\right) < 0$) and $R/\lambda = 1/2$
 for dielectrics ($\Re\left(\chi\right) > 0$). Unless additional
 constraints are included, findings for dielectric materials limited
 to small domains, and findings for metals in large domains, exhibit
 substantially larger disagreement.} 
 \label{fig4_farfieldBounds}
\end{figure*}

Exploiting this relation and the incoherent nature of the 
fluctuations, the net emitted power may be expressed as a sum over 
independent radiative channels. 
Generally, the instantaneous power emitted by a
current source is
\begin{equation}
 P_{rad} =
  -\int_{\mathsf{R}}
  \mathbf{J}(\mathbf{x}, t) \cdot \mathbf{E}(\mathbf{x}, t)
 \label{eqWorkDone}
\end{equation}
where the minus sign results from the convention of emitted power.
Switching to the spectral domain~\footnote{The convention used here is
 $\mathbf{J}(\mathbf{x}, t) = \int_{-\infty}^{\infty}
 d\omega~e^{-i\omega t} \mathbf{J}(\mathbf{x}, \omega)$.}, and
assuming that the collection of fluctuating dipolar sources
distributed throughout the body satisfy
Eq.~\eqref{fluctuationDissipation}, the thermal power radiated by a
body held at a constant temperature $T$
(c.f.~\cite{kruger2017RevNonEq, kruger2012trace,molesky2019bounds}) is
given by
\begin{align}
 \{P_{rad}\}_{T}
 &= \int\limits_{0}^{\infty}d\omega~\Pi(\omega, T) 
 ~\underbrace{\frac{2}{\pi} \text{Tr}\, [
  \mathbb{G}_{0}^{\mathsf{A}}(\mathbb{T}^{\mathsf{A}} -
  \mathbb{T}^{\dagger}
  \mathbb{G}_{0}^{\mathsf{A}}\mathbb{T})]}_{\Phi(\omega)}
\label{eqHandPhi}
\end{align}
where $\Pi(\omega,T) =\frac{\hslash\omega}{\exp(\hslash\omega/
 \left(k_{B}T\right)) - 1}$ is the Planck thermal occupation 
function, and $\Phi(\omega)$ the corresponding angle-integrated
spectral transfer function (absorption or emission) of the body; 
the Tr symbol denotes a trace over both the position and polarization 
indices of the dipole sources, i.e. the complete set of indices of 
the enclosed operators.

In the breakup of Eq.~\eqref{eqHandPhi}, 
the $\mathbb{T}^{\mathsf{A}}-\mathbb{T}^{\dagger} 
\mathbb{G}_{0}^{\mathsf{A}}\mathbb{T}$ term contained in 
$\Phi$ constitutes an algebraic description of
absorption, Eq.~\eqref{matLoss}, expressed as the subtraction of
radiated power $\mathbb{T}^{\dagger}
\mathbb{G}_{0}^{\mathsf{A}}\mathbb{T}$, Eq.~\eqref{radLoss}, from the
total extracted (extinction) power $\mathbb{T}^{\mathsf{A}}$. 
This association is no accident: as a consequence of reciprocity, 
evaluating the trace over a (delocalized) basis of waves incident on
the body changes the interpretation of $\Phi$ from the net emitted power 
due to dipolar sources within the object (thermal emission) to 
the net power absorbed in the body due to incident plane waves 
(angle-integrated absorption), but the algebraic 
form of $\Phi$ remains unaltered. 
Because $\mathbb{G}_{0}^{\mathsf{A}}$ describes how outgoing
radiation carries power away from an object into the
surrounding environment~\cite{landau2013statistical,
molesky2019bounds}, a natural basis in which to evaluate the trace is
the eigenmode expansion
\begin{align}
 \mathbb{G}_{0}^{\mathsf{A}} = 
 \sum_{n} \rho_{n}\left|\mathbf{Q}_{n}\right>
 \left<\mathbf{Q}_{n}\right|,
\label{asymGexpansion}
\end{align}
with each of the radiative coefficients $\rho_{n}$ nonnegative by
passivity. 
Setting $\left|\mathbf{T}_{n}\right> = \mathbb{T}\left
|\mathbf{Q}_{n}\right>$~\footnote{Up to a constant,
 $\left|\mathbf{T}_{n}\right>$ is the polarization current resulting
 from the $n$-th radiative mode.}, $\Phi$ becomes
\begin{equation}
 \Phi = \frac{2}{\pi}\sum_{n}\rho_{n}\left(\Im\left[
 \left<\mathbf{Q}_{n}|\mathbf{T}_{n}\right>\right] - 
 \left<\mathbf{T}_{n}\right|\mathbb{G}_{0}^{\mathsf{A}}
 \left|\mathbf{T}_{n}\right>\right).
 \label{phiSum}
\end{equation}
Even without the imposition of a single constraint, the form of 
Eq.~\eqref{phiSum} places fairly strong restrictions on the extent 
to which the net absorption (resp. emission) cross section of an object 
can be enhanced compared to its geometric cross 
section~\cite{molesky2019bounds}. 
In order to optimize absorption it is clear from Eq.~\eqref{phiSum} that
each radiative mode must generate a strong polarization: $\Im\left[
\left<\mathbf{Q}_{n}|\mathbf{T}_{n}\right>\right]$ is the extracted 
power. 
However, the generation of these currents necessarily leads to
radiative losses, $\left<\mathbf{T}_{n}\right|
\mathbb{G}_{0}^{\mathsf{A}}\left|\mathbf{T}_{n}\right>$, which grow
relatively in strength as the size of the domain increases through the 
growth of the $\rho_{n}$ radiative coupling 
coefficients~\cite{molesky2019bounds,venkataram2020fundamental}. 

As a first example of optimization bounds, we analyze the maximization
of $\Phi$ subject to the constraint that real power is conserved:
\begin{align}
 &\max_{\{\left|\mathbf{T}_{n}\right>\in\Omega\}}~
 \frac{2}{\pi}\sum_{n}\rho_{n}\left(\Im\left[
 \left<\mathbf{Q}_{n}|\mathbf{T}_{n}\right>\right] - 
 \left<\mathbf{T}_{n}\right|
 \mathbb{G}_{0}^{\mathsf{A}}\left|\mathbf{T}_{n}\right>
 \right)
 \nonumber \\
 &\text{such that}~\forall n
 \nonumber \\
 &\Im\left[\left<\mathbf{Q}_{n}|\mathbf{T}_{n}\right>\right] - 
 \left<\mathbf{T}_{n}\right|
 \left(\bm{\chi}^{-1\dagger}
 -\mathbb{G}_{0}^{\dagger}\right)^{\mathsf{A}}
 \left|\mathbf{T}_{n}\right> = 0,
 \label{realPowerProb}
\end{align}
effectively the simplest version of Eq.~\eqref{qcqp}. 
Due to the form of $\Phi$, the only difference between the objective and
constraint in Eq.~\eqref{realPowerProb} is the material loss term
$\lVert\mathbf{T}_{n}\rVert^{2}/\zeta$, with the factor $\zeta\equiv
\lVert\chi\rVert^{2}/\Im\left[\chi\right]$ quantifying the maximum
magnitude that the polarization current density can achieve relative
to the incident electric field~\cite{miller2015shape}. Simply, to
maintain equilibrium, the net (integrated) power extracted by the
object at each frequency must be perfectly balanced by the sum of two
possible loss mechanisms: the absorption of power into material
degrees of freedom ($\bm{\chi}^{-1\dagger\mathsf{A}}$), here
considered to be an infinitely large thermal bath~\cite{xu2019review},
and power re-radiated (scattered or reflected) back into the ambient
environment ($\mathbb{G}_{0}^{\mathsf{A}}$).

Applying the relaxation of Lagrange duality (c.f. 
Refs.~\cite{boyd2004convex,beck2006strong,angeris2019computational,
molesky2020t,angeris2021heuristic}), the optimal objective value of 
Eq.~\eqref{realPowerProb} can be expressed as
\begin{widetext}
\begin{align}
 \Phi_{opt}
 = \frac{1}{2\pi}\sum_{n}
 \begin{cases}
   ~1 &\zeta\geq \frac{1}{\rho_{n} }\\
   \frac{4\zeta\rho_{n}}{(1 + \zeta\rho_{n})^{2}}
   &\zeta < \frac{1}{\rho_{n}}\\
 \end{cases}
 \leftrightarrow
  \begin{cases}
   ~1 &\tau_{m,n}\geq \tau_{r,n}\\
   \frac{4\tau_{m,n}\tau_{r,n}}{(\tau_{r,n} + 
   \tau_{m,n})^{2}} 
   &\tau_{m,n} < \tau_{r,n}\\
 \end{cases},
 \label{realPowerSol}
\end{align}
\end{widetext}
with $\leftrightarrow$ marking associations,
$\tau_{r,n}\leftrightarrow 1/\rho_{n}$ and $\tau_{m,n} \leftrightarrow
\zeta$, with a coupled-mode analysis, Box.~1. The surprising
simplicity of $\Phi_{opt}$ as arising from a sum over
independent channel contributions follows from the fact that, absent
other scattering constraints (beside real power conservation), the
optimal bound polarization currents end up becoming diagonal in the
basis of radiation states (the eigenbasis of
$\mathbb{G}_{0}^{\mathsf{A}}$), $\ket{\vec{T}_{n}} \approx 
c_{n}\left|\mathbf{Q}_{n}\right>$ (neglecting non-radiative terms that
yield minor modifications), with the maximum bound polarization
response $\lVert c_{n}\rVert \leq \min\left\{\frac{1}{2\rho_{n}},
\zeta\right\}$ in each channel.

A trio of plots of $\Phi_{opt}$, bounding angle-integrated absorption
(equivalently emission) for bodies enclosed in a spherical ball of
radius $R$, are shown in Fig.~\ref{fig4_farfieldBounds}. Beyond the
exact quantitative predictions appearing in \textbf{a}, and the
excellent agreement with computationally designed structures appearing
in \textbf{b} (see Refs.~\cite{polimeridis2015fluctuating,
 molesky2019bounds} for details), it is seen that the mere
conservation of net real power is sufficient for
Eq.~\eqref{realPowerProb} to inherently reproduce fundamental
quasi-static and blackbody behavior. In the limit of a small design
volume, $\zeta\rho_{n} \ll 1$ (resp. $\tau_{m,n} \ll \tau_{r,n}$) for
all $n$, $\Phi_{opt}$ is seen to exhibit a volumetric scaling
consistent with the assumption that the magnitude of all generated
polarization currents can grow as large as material loss allows: as
the volume grows, so does the available power in each channel, 
and hence so should the polarization response. However, due to
the necessary coupling of these currents with radiative states,
volumetric growth cannot persist indefinitely. Eventually, in each
index of Eq.~\eqref{phiSum}, growth in $\rho_{n}$ (resp. decay in
$\tau_{r,n}$) causes radiative losses to discordantly overwhelm net
extracted power if the magnitude of $\left|\mathbf{T}_{n}\right>$
(resp. the material lifetime $\tau_{m,n}$) becomes too large, leading
the associated channel (index) to enter the saturation condition of
Eq.~\eqref{realPowerSol}, visible in Fig.~\ref{fig4_farfieldBounds} as
the onset of steps. As an increasing number of channels saturate,
volumetric scaling begins to asymptotically transition to area
scaling, regardless of the supposed value of $\zeta$. Directly, the
black-body limit equating absorption and geometric cross sections
appears out of Eq.~\eqref{realPowerProb}, irrespective of 
assumed material properties.

The behaviors of the $\rho_{n}$ radiative expansion coefficients also
have interesting implications for antenna
design~\cite{capek2017minimization,gustafsson2019tradeoff}. For any
excitation contained within $\Omega$, the largest radiative
coefficient $\rho_{n}$ in Eq.~\eqref{asymGexpansion} sets a lower
bound on the radiative lifetime, $\tau_{r}$, and consequently, a lower
bound on the radiative quality factor $Q_{r} = \omega\tau_{r}/2$. In
the limit of a deeply subwavelength design volume, the largest
$\rho_{n}$ (the dipole coefficient) scales $\propto
\left(k_{0}R\right)^{3}$, setting a lower bound on the radiative
quality factor $\propto\left(k_{0}R\right)^{-3}$~\cite{chu1948physical,
 harrington1960effect,mclean1996re}. This lower bound on $Q_{r}$,
known as the Chu limit ($Q_\mathrm{rad} \geq \frac{1}{k_{0}R} +
\frac{1}{(k_{0}R)^3}$~\cite{mclean1996re}), correspondingly sets an
upper limit on antenna bandwidth that restricts processing
speed~\cite{capek2016optimal}.

Finally, as suggested above, the sum expression for $\Phi_{opt}$ given
by Eq.~\eqref{realPowerSol} is completely analogous to a modal
description of absorption by non-interacting excitations and, as one
might expect, under this analogy the channel saturation condition of
$\zeta\rho_{n} = 1$ is exactly the rate-matching condition $\tau_{r,n}
= \tau_{m,n}$ between radiative and material lifetimes. Consequently,
it is accurate to interpret Eq.~\eqref{realPowerSol} as a model
wherein an idealized object independently extracts power from each of
the radiative (multipole) states in a physically optimal way, with
$\rho_{n}$ and $\zeta$ setting limits on the associated coupled-mode
radiative and dissipation rates for every channel. As discussed in
greater detail below, this perspective implicitly assumes that
resonant response is achievable in each individual channel, and that
attaining resonant response in one channel has no implications for any
other. Neither assumption is typically true in practice.
\\ 
\begin{figure*}[t]
\centering 
 \includegraphics[width=1.0\textwidth]{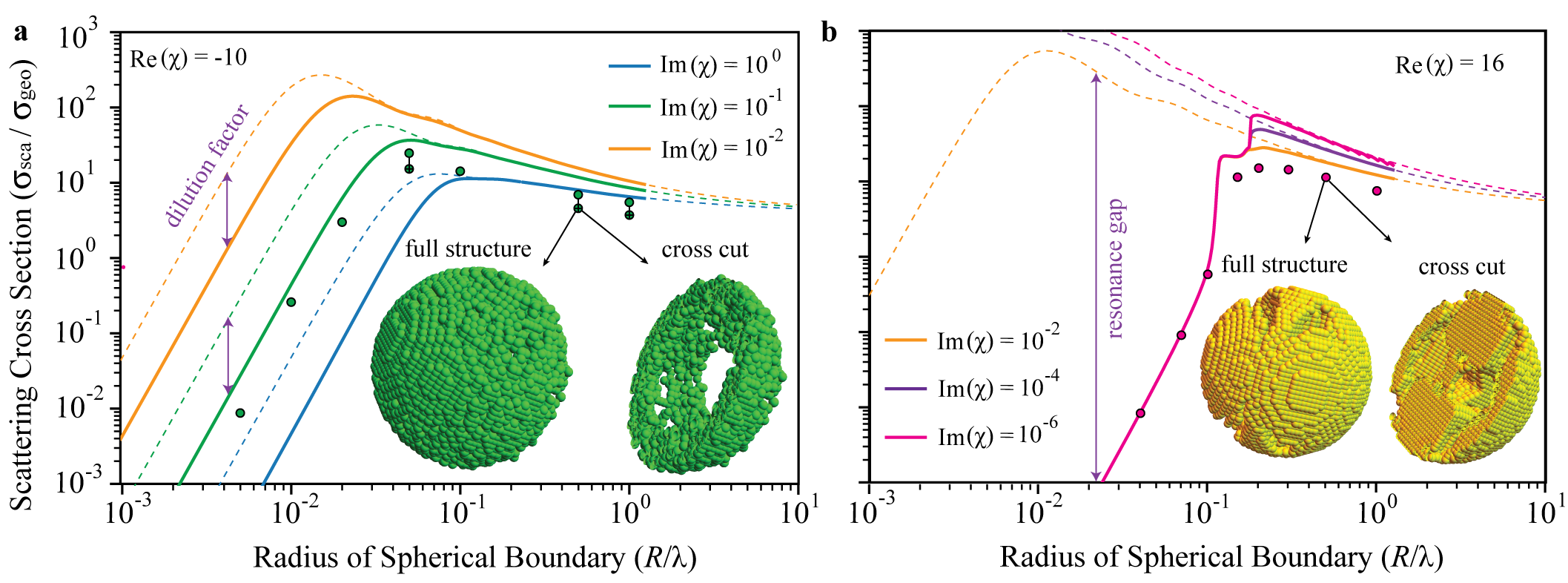}
 \caption{\textbf{Bounds on scattering cross
 sections based on the conservation of total power.} 
 Panels \textbf{a} and \textbf{b}, taken from
 Ref.~\cite{molesky2020t}, depict the results of
 Eq.~\eqref{reactPowerProb} as a function of the size (radius $R/
 \lambda$) of a spherically bounded design volume for two
 representative values of $\Re\left[\chi\left(\omega\right)\right]$ 
 near optical frequencies (a near plasmonic metal 
 $\textbf{a}$ and a strong dielectric $\textbf{b}$.) 
 Dashed lines result from only imposing the conservation of real power; 
 solid lines result from additionally imposing the conservation of 
 reactive power. 
 The dots appearing in both panels mark scattering cross sections
 achieved in actual geometries discovered by numeric (inverse) design, 
 for $\chi = -10 + i 10^{-1}$ and $\chi = 16 + i 10^{-6}$ respectively. 
 For the metal structures in \textbf{a}, aligned cross-hatched dots 
 result from binarizing the discovered permittivity profiles, 
 which are otherwise allowed to take on ``gray-scale'' 
 values~\cite{christiansen2021inverse}. 
 Two sample structures are shown as insets, with the planewave incident 
 from the more solid side of both designs, from the left in \textbf{a}, 
 from the right in \textbf{b}, and aligned along the left-right symmetry 
 axis. 
 Again, in the limit that $R/\lambda\gg 1$ or $R/\lambda \ll 1$ the 
 calculated bounds approach the scaling predictions of ray optics 
 (geometric cross sections) and quasi-static scattering ($\propto V^{2}$), 
 irrespective of the electric susceptibility.
 The inclusion of reactive power conservation for strong metals 
 ($\Re\left[\chi\left(\omega\right)\right] \ll -3$) confined 
 to small design volumes ($R/\lambda \ll 1$) shows that the 
 structuring needed to achieve resonant scattering away 
 from the plasmon condition of $\Re\left[\chi\left(\omega\right)\right] 
 = -3$ reduces achievable material scaling characteristics---the 
 appearance of ``dilution factors''.
 The of effect total power conservation on obtained limits for dielectric 
 materials confined to subwavelength domains is more radical---causing the 
 appearance of ``resonance gaps''---leading to the general conclusion that 
 scattering cross section enhancements surpassing $\approx 200$ should not 
 be expected for near optical frequencies.}
 \label{fig5_SctLimits}
\end{figure*}
\\ \\ 
\textbf{Total power conservation}---Given the remarkable agreement
observed in Fig.~\ref{fig4_farfieldBounds}\textbf{b} between the
absorption characteristics of structures obtained by computational
techniques and the associated bounds, it is natural to wonder
whether related conclusions can be drawn for all basic scattering
quantities; namely, to what extent are the results of current inverse
methods explained by the basic necessity of conserving power?

Recalling that the power scattered from an initial electric field
$|\mathbf{E}^{i}\rangle$ at a single frequency $\omega$
is~\cite{tsang2004scattering,molesky2020t}
\begin{align}
  P_{sct} 
  &=
  \frac{k_{0}}{2Z}\left<\mathbf{E}^{i}\right|\left[
 \mathbb{T}^{\mathsf{A}} - 
 \mathbb{T}^{\dagger}\left(\mathbb{V}^{-1\dagger}
 \right)^{\mathsf{A}}
 \mathbb{T}\right]\left|\mathbf{E}^{i}\right>,
 \label{PsctToperator}
\end{align}
the problem of maximizing the scattering cross section of an object
contained within a design volume $\Omega$ becomes the optimization
statement
\begin{align}
 &\max_{\left|\mathbf{T}\right>\in\Omega}~
 \frac{k_{0}}{2Z}
 \left[\Im\left(\left<\mathbf{E}^{i}|\mathbf{T}\right>\right)
 -
 \left<\mathbf{T}\right|
 \bm{\chi}^{-1\dagger\mathsf{A}}\left|
 \mathbf{T}\right>\right]
 \nonumber \\
 &\text{such that}
 \nonumber \\
 &\Im\left(\left<\textbf{E}^{i}|\textbf{T}\right>\right) - 
 \left<\textbf{T}\right|
 \left(\bm{\chi}^{-1\dagger}
 -\mathbb{G}_{0}^{\dagger}\right)^{\mathsf{A}}
 \left|\textbf{T}\right> = 0, \nonumber \\
 &\Re\left(\left<\textbf{E}^{i}|\textbf{T}\right>\right) - 
 \left<\textbf{T}\right|
 \left(\bm{\chi}^{-1\dagger}
 -\mathbb{G}_{0}^{\dagger}\right)^{\mathsf{S}}
 \left|\textbf{T}\right> = 0,
 \label{reactPowerProb}
\end{align}
where $|\mathbf{E}^{i}\rangle$ is the electric field of an incident
plane wave and, as before, $\left|\mathbf{T}\right> =
\mathbb{T}\left|\mathbf{E}^{i}\right>$ so that
$\left(-ik_{0}/Z\right)\left|\mathbf{T}\right> = \left|\mathbf{J}\right>$ 
is the resulting electric polarization current density in the object. 
If only the conservation of real power is imposed, the solution of
Eq.~\eqref{reactPowerProb} closely mirrors the coupled-mode expression
for integrated emission given by Eq.~\eqref{realPowerSol}. However, as
is clear from a comparison of the solid (enforcing total power
conservation) and dashed (enforcing only real power conservation)
lines of Fig.~\ref{fig5_SctLimits} (particularly \textbf{b}), the
inclusion of the global reactive power constraint, the final line of
Eq.~\eqref{reactPowerProb}, leads to substantially tighter
limits~\footnote{Similar limit tightening also occurs when the
 conservation of reactive power is enforced for thermal radiation as
 applied to dielectric materials and subwavelength design domains.},
and under the imposition of this additional constraint, barring the
single channel asymptotic examined in Box.~1, the solution of
Eq.~\eqref{reactPowerProb} does not have a simple semi-analytic
form. The physical mechanism underlying these differences is the
appearance of phase information. Paralleling the well-known response
characteristics of a simple harmonic oscillator, when only the
conservation of real power is enforced there is a bound on the
relative \emph{magnitude} of $\left|\textbf{T}\right>$ set by material
absorption and radiative emission. All information about the relative
phase offset of the response, partially set by
$\Re\left[\chi\left(\omega \right)\right]$, is ignored.
Once the need to conserve reactive power is taken into account,
physical restrictions are placed on both the relative \emph{magnitude}
and \emph{phase} of the response~\cite{jackson1999classical,
 gustafsson2020upper,molesky2020t}. These restrictions can either
limit, or even exclude, resonant response~\footnote{Here, by resonant
 response, we mean that, under sufficiently small changes to the
 material properties of a design, there is a field observable that
 scales roughly as $\propto \zeta =
 \lVert\chi\left(\omega\right)\rVert^{2}/\Im\left[
  \chi\left(\omega\right)\right]$.} in certain situations. 
\newpage
\onecolumngrid 
\noindent
\textbf{Box~1: Single-channel asymptotics (antenna cross sections)} 
\\
\rule{1.0\textwidth}{1.0 pt}
\\ 
\twocolumngrid 

\begin{figure}[H]
\centering 
 \includegraphics[width=0.5\textwidth]{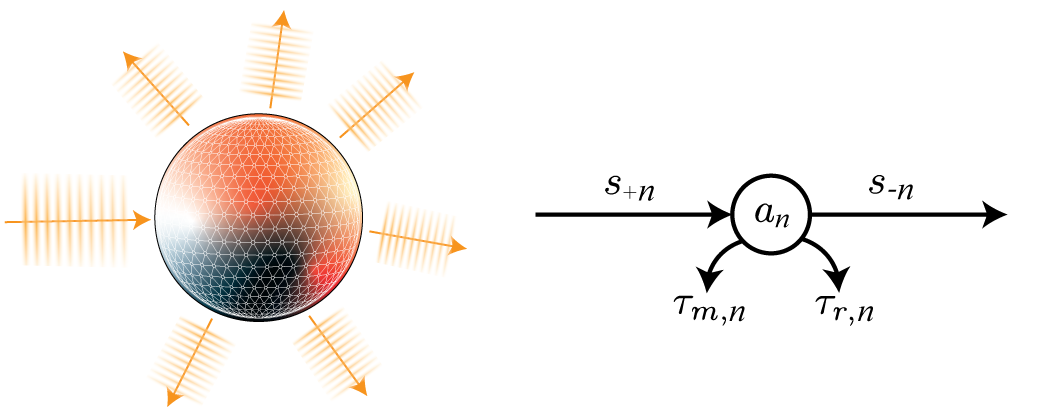} 
 \caption{\textbf{Schematic
 of coupled-mode theory.} The figure, adapted from
 Ref.~\cite{hamam2007coupled}, illustrates the basic coupled-mode
 analysis of resonant scattering discussed within. To each mode
 (channel) $n$ of some scattering object, represented by the
 amplitude $a_{n}$, there are decay channels associated with
 radiative emission, the radiative lifetime $\tau_{r,n}$, and
 material absorption, the material lifetime $\tau_{m,n}$. Whenever
 an incident field overlaps with the mode indexed by $n$, power is
 transferred from the incident field, $s_{+n}$, into the mode
 $a_{n}$, defining ``ports'' into and out of the channel. Supposing
 that each mode $n$ is orthogonal to every other (e.g., if the modes
 are the radiative solutions of a small domain) the scattering cross
 section of the object can then be defined by summing over
 $n$.} \label{fig6_CMT}
\end{figure}
As a means of gaining further insight, it is useful to examine the
solutions of Eq.~\eqref{reactPowerProb} in the single-channel
asymptotic (quasistatic) regime of small domain sizes $R\ll\lambda$
through the lens of modal decompositions and coupled-mode theory---a
type of time-dependent perturbation theory (also known as
Breit--Wigner scattering theory~\cite{landau2013quantum}) born out of
the assumption that the optical response of an object may be described
as a sum of weakly interacting resonant modes coupled to one another
and/or incident fields via radiative channels~\cite{haus1991coupled}.
Following the work of Hamam et al.~\cite{hamam2007coupled} along the
lines described below Fig.~\ref{fig6_CMT}, through a combination of
real power conservation and time-reversal symmetry, it can be shown
that the amplitude $a_{n}$ of each such mode obeys the equation
$da_{n}/dt = \left[-i\omega_{n} - \left( \frac{1}{\tau_{r,n}} +
 \frac{1}{\tau_{m,n}}\right)\right]~a_{n} +
\sqrt{\frac{2}{\tau_{r,n}}}~s_{+n},$ where $\omega_{n}$ denotes the
frequency of oscillation of mode $n$. Solving for
$a_{n}\left(t\right)$, using the scattering cross section formula
$P_{sct}/I_{0} = 2\lVert a_{n}\rVert^{2}/ \left(\tau_{m,n} \lVert
a_{n}\rVert^{2}\right)$, as applied to a $\left(2n+1\right)$-fold
degenerate state (e.g. the total angular momentum number $n$ of
radiative solutions in spherical coordinates), one finds
\vspace{-6pt}
\begin{equation}
 \frac{\sigma_{sct,n}}{\sigma_{geo}}
 = 2\left(2n+1\right)
 \frac{\left(1/\tau_{r,n}\right)^{2}}{\left(\omega -
 \omega_{n}\right)^{2} + \left(\frac{1}{\tau_{m,n}} + 
 \frac{1}{\tau_{r,n}}\right)^{2}}
 \left(\frac{\lambda}{2\pi R}\right)^{2},
 \label{cmScatCS}
\end{equation}
with $\sigma_{geo} = \pi R^{2}$ denoting the geometric cross section
of a sphere of radius $R$. Intuitively, an expression like
Eq.~\eqref{cmScatCS} can also be considered as a bound, set by the
conservation of real power, on maximum scattering: assuming that each
channel can be designed to have the same resonance frequency, the
parametric relations imply a set of rate matching conditions (e.g.,
$\tau_{r,n} = \tau_{m,n}$ in Eq.~\eqref{cmScatCS}) for maximizing each
channel's contribution to scattering at some wavelength
$\lambda$. While providing a predictive and conceptual model of
scattering, with meaningful insights into a wide range of applications
c.f. Refs.~\cite{geyi2003physical,
 kwon2009optimal,verslegers2010temporal,ruan2012temporal,munday2012light,
 liberal2014upper,jia2015theory,nordebo2017physical}, such modal
descriptions face significant challenges in setting quantitative
limits. Explicitly, it is seldom clear (1) how many channels should
be considered, i.e. where the sum should be cut-off to avoid
divergence without being overly
restrictive~\cite{pendry1999radiative,ben-abdallahFundamental2010};
(2) what range of parameter values are possible for each channel; and
(3) to what level the parameters are connected; i.e. does material
structuring allow for independent parameter tuning.

When Eq.~\eqref{reactPowerProb} is considered on a sufficiently small
ball, the decomposition of an incident planewave source into radiative
multipoles approximately terminates at the ($n=1$) dipole fields.
Under this quasistatic condition, Eq.~\eqref{reactPowerProb} can be
solved semi-analytically~\cite{molesky2020t}, leading to the
single-channel $R\ll\lambda$ asymptotics
\begin{equation}
 \frac{\sigma_{sct}^{opt}}{\sigma_{geo}} 
 \leq\frac{3}{2}
 \begin{cases}
  \frac{\rho_{1}^{2}}{\left(\frac{1}{3}
   + 
  \frac{\Re\left(\chi\right)}{~\lVert\chi\rVert^{2}}\right)^{2}
  +\left(\rho_{1} + 
  \frac{\Im\left(\chi\right)}{~\lVert\chi\rVert^{2}}\right)^{2}}
  &\Re\left(\chi\right) > - 3 \\
  \frac{\left[\rho_{1}\lVert\chi\rVert/\Re\left(\chi\right)
  \right]^{2}}{\left[\rho_{1} + \frac{\Im\left(\chi
  \right)}{3\lVert\Re\left(\chi\right)\rVert}\right]^{2}}
  & \Re\left(\chi\right) \leq -3
 \end{cases}
 \label{singleChannelSctRective}
\end{equation}
where $\rho_{1} = \frac{2}{9}\left(2\pi R /\lambda\right)^{3}$ 
is the radiative coefficient of the three-fold degenerate dipole 
channel, Eq.~\eqref{asymGexpansion}. 
Up to a factor of $4$, which as originally explained in 
Ref.~\cite{miller2016fundamental} occurs because optimal scattering 
simultaneously implies optimal absorption, there is again a clear 
association between Eq.~\eqref{singleChannelSctRective} and 
the coupled-mode theory expression given by Eq.~\eqref{cmScatCS}. 
So long as $\Im\left[\chi\left(\omega\right)\right]$ is small compared to 
the absolute value of $\Re\left[\chi\left(\omega\right)\right]$, 
Eq.~\eqref{singleChannelSctRective} is consistently interpreted 
in terms of coupled mode parameters as
\begin{align}
\omega - \omega_{1} &\leftrightarrow 1/\Re\left(\chi\right) - 
\left(-1/3\right) 
~~~~\tau_{r,1} \leftrightarrow 1/\rho_{1} \nonumber \\
\tau_{m,1} &= 
\begin{cases}
\lVert \chi\rVert^{2} / \Im\left(\chi\right), & 
\Re\left(\chi\right) > -3 \\
3 \,\lVert \Re\left(\chi\right)\rVert/ \Im\left(\chi\right), 
& \Re\left(\chi\right) \leq -3
\end{cases}
\end{align}
With regards to optimal response, Eq.~\eqref{singleChannelSctRective}
shows that ``truly'' resonant response
($\Im\left(\chi\right)/\lVert\chi\rVert^{2}$ scaling) is only possible
when $\Re\left[\chi\left(\omega\right)\right] = -3$, the localized
plasmon resonance of a spherical
nanoparticle~\cite{novotny2012principles}. For all other material
choices, Eq.~\eqref{singleChannelSctRective} implies that, when a
resonance is possible, some amount of material structuring is needed
to achieve a maximized scattering cross section, and, in
undertaking this structuring, the potential for field enhancement is
reduced. In the case of dielectrics and weak metals
($\Re\left(\chi\right) > -3$) ``resonant'' scattering is simply not
possible. In the case of plasmonic metals ($\Re\left(\chi\right) <
-3$), resonant response is only possible through a ``dilution'' of the
effective electric susceptibility to the plasmon condition,
Fig.~\ref{fig5_SctLimits}\textbf{a}; see Ref.~\cite{molesky2020t} for
additional details.
\onecolumngrid 
\vspace{-2pt}
\noindent
\rule{1.0\textwidth}{1.0 pt}
\\
\twocolumngrid
Most notably, as confirmed by Fig.~\ref{fig5_SctLimits}\textbf{b}, when
confined to a spherical subwavelength domain, the largest possible
scattering cross section that can be achieved by structuring a
dielectric material is \emph{exactly} the $\propto V^{2}$ Rayleigh
scattering of a ball~\cite{hulst1981light} (see Box.~1 and
Ref.~\cite{molesky2020t} for further details).

The inclusion of reactive power also implies drastic alterations to
the mathematical model and interpretation of
Eq.~\eqref{reactPowerProb} as compared to Eq.~\eqref{realPowerProb}.
Precisely, the sum form of $\Phi_{opt}$ given by
Eq.~\eqref{realPowerSol} arises because both the objective and
constraint of Eq.~\eqref{realPowerProb} are diagonalized in an
eigenbasis of $\mathbb{G}_{0}^{\mathsf{A}}$~\footnote{It 
should be kept in mind that this description often accurately 
anticipates the performance of structures discovered by 
computational methods.}. 
When further physical constraints are imposed (e.g., global 
reactive power or the local constraints introduced in the next 
set of examples) simultaneous diagonalization is rarely possible. 
In part indicating the richness of
the physics being described, the symmetric and anti-symmetric
components of the Green's function, $\mathbb{G}_{0}^{\mathsf{S}}$ and
$\mathbb{G}_{0}^{\mathsf{A}}$, do not share a common
eigenbasis~\cite{harrington1972characteristic,
 molesky2020hierarchical}. Hence, whenever response characteristics
are not dominated by the conservation of real power, it is generally
not possible to describe scattering phenomena in terms of an orthogonal
basis of weakly interacting modes. Once both aspects of the Green's
function are included, radiative channels are mixed both among
themselves and with non-radiative states. Relatedly, although the
content of Box.~1 stands an exception, there is typically no simple
way to use optimization bounds as means of doing parameter extraction
for coupled mode theory. 
\\ \\ 
\textbf{Local constraints}---Extrapolating
from the last two examples, enforcing that total power must be
conserved is found to capture almost all relevant physical 
effects that limit achievable scattering characteristics for 
propagating waves---far-field applications 
like maximizing planewave absorption, thermal 
radiation~\cite{molesky2019bounds}, and scattering cross 
sections~\cite{molesky2020t,gustafsson2020upper,kuang2020maximal,
trivedi2020bounds}. 
However, when applied to objectives governed by rapidly 
varying (e.g., evanescent) fields or multiple length scales, 
these coarse characterizations of the integral-wave relations 
may miss important aspects of the problem, leading to bounds 
with little connection to reality~\cite{molesky2020hierarchical,
kuang2020computational}.
To remedy this issue, additional physics in the form of localized
constraints incorporating higher spatial resolution must be 
included. 
Specifically, contrasting Eq.~\eqref{reactPowerProb} with the system of
equations that must be solved in order to approximate Maxwell's
equations numerically, it is clear that imposing global power
conservation (a total of two constraints) cannot possibly capture all
relevant wave features: the use of $\mathbb{P} = \mathbb{I}_{_\Omega}$
in Eq.~\eqref{braKetConstraints} only guarantees that scattering
theory is true \emph{on average} over the volume of the
domain. At any spatial point, the solution resulting from an
optimization problem statement like Eq.~\eqref{reactPowerProb} may 
violate Eq.~\eqref{tOpt1}, so long as these violations cancel each
other when integrated over the entire scattering domain.

\begin{figure*}[t]
\centering
\includegraphics[width=1.0\textwidth]{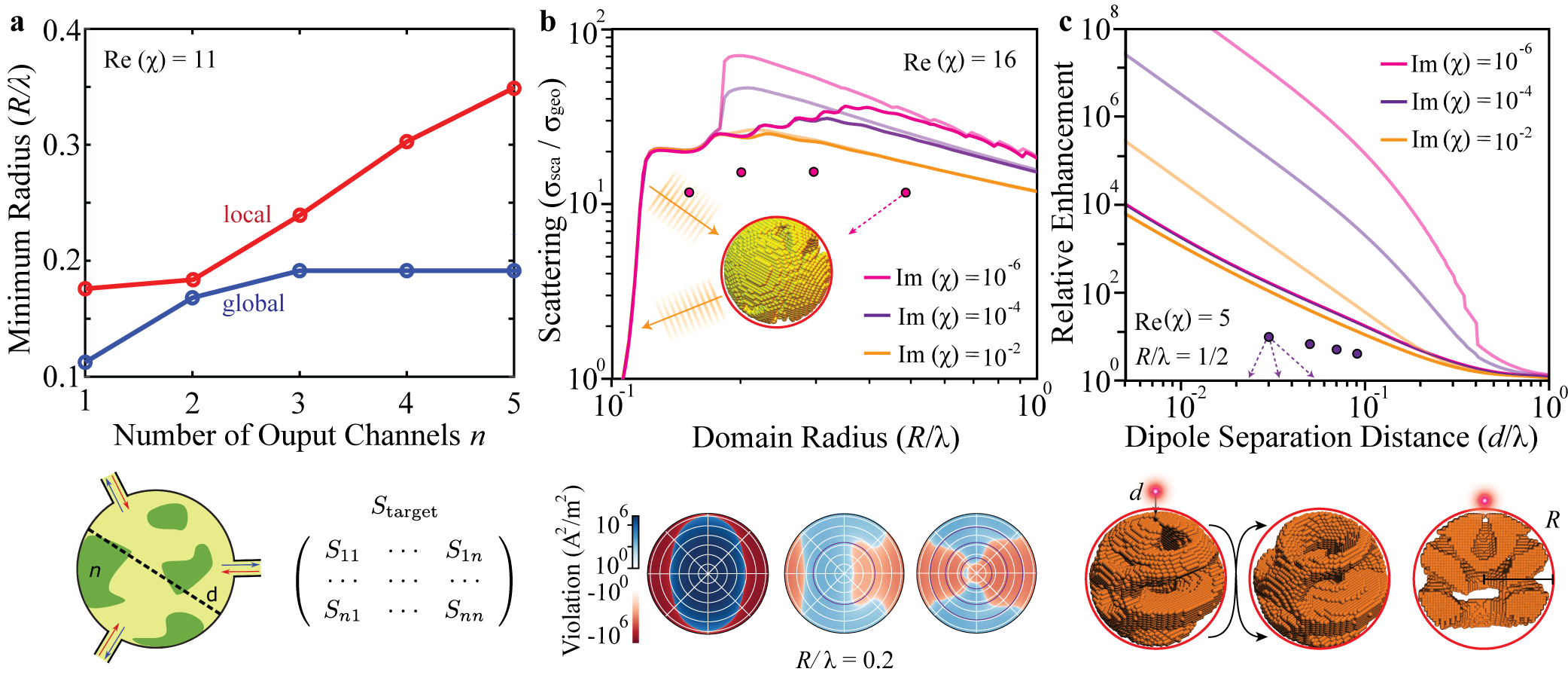}
\vspace{-10 pt}
\caption{\textbf{Impact of localized constraints.} 
The figure shows upper bounds under the enforcement of local 
constraints on \textbf{a} the minimum radius $R$ of a cylindrical 
design domain necessary to encode a ``power-splitting'' 
$\mathbb{S}$ matrix, evenly distributing power from a single 
plane-wave input into $2n+1$ outgoing radiative channels 
(Ref.~\cite{kuang2020computational}); \textbf{b} plane-wave 
scattering as described in the previous tutorial 
(Ref.~\cite{molesky2020hierarchical}); and \textbf{c} radiative 
Purcell enhancement for an electric dipole source in the near 
field of an object (Ref.~\cite{molesky2020hierarchical}). 
Lighter-colored lines in \textbf{b} and \textbf{c} result from 
imposing the global conservation of real and reactive power. 
Like-colored darker lines are obtained by enforcing local
conservation constraints over radial shell subdomains (varying 
distribution with application). 
As in previous plots, dots in \textbf{b} and \textbf{c} mark 
actual enhancement values achieved by structures discovered 
by computational methods. 
Views of one such design are given below \textbf{c}. 
The logarithmic color maps included below \textbf{b} indicate 
the local violation of the conservation of reactive power for 
$1,2$, and $4$ evenly spaced shell subregion constraints. 
The schematic shown below \textbf{a} conceptualizes the engineering 
of a device to realize a particular scattering matrix. 
In all three panels, additional constraints lead to increasingly 
realistic field features and limit values.}
\label{fig7_localConstraints}
\end{figure*}

Owing to the nature of the relaxation techniques commonly used to
solve optimization bounds (c.f. Refs.~\cite{molesky2020hierarchical,
angeris2021heuristic}), the spatial oscillation of such local 
violations of ``true'' physics tend to track the spatial oscillations 
of the incident field, as seen in 
Fig.~\ref{fig7_localConstraints}\textbf{b}, and accordingly, by 
imposing localized projections 
($\mathbb{P} = \mathbb{I}_\mathsf{P}$ where $\mathsf{P} \subset \Omega$) 
in Eq.~\eqref{braKetConstraints} targeting a specific region of 
violation, it is usually possible to guide optimization limits 
towards increasingly physical characteristics: the vacuum Green's 
function $\mathbb{G}_{0}$ does not propagate all information equally, 
but rather blurs rapid field fluctuations as the point of observation 
moves away from the source.
As such, for each design problem, there are characteristic lengths
below which differences between field solutions have no pragmatic
relevance. If independent constraints enforce ``averaged'' physics on
a grid finer than the smallest of these length scales, then the
solution of Eq.~\eqref{qcqp} should differ little from what is
realizable in practice~\cite{angeris2021convex}. Directly, the number
and distribution of local constraints can be viewed as “tuning knobs”
enforcing physics at the expense of computational complexity.

An exemplification of these ideas is illustrated by the 
minimum-radius limits on a two-dimensional power splitter, 
distributing power from an incident wave
equally between $2n+1$ cylindrical wave channels, depicted in
Fig.~\ref{fig7_localConstraints}\textbf{a}
(Ref.~\cite{kuang2020computational}). In asserting only the
conservation of global power, the minimum diameter asymptotes to
$\sim\lambda/3$, a size at which it begins to become possible for a
non-physical response oscillating near $\lambda/\sqrt{\Re(\chi) + 1}$
to satisfy global power conservation while maintaining large local
violations. Opposingly, when local constraints are added, this
asymptote disappears and the required radius (suggesting increasing
device complexity) begins to grow steadily with the number of power
divisions desired. In Fig~\ref{fig7_localConstraints}\textbf{b}, the
strict use of global constraints similarly suggests that shortly after
attaining a resonance criteria of $R\gtrsim \lambda /5$, by
structuring a material with $\Re(\chi) = 16$ within a spherical ball of
radius $R$, it is possible to enhance the scattering cross section
inversely proportional to material loss ($\Im(\chi)$). No such scaling
is found in computationally synthesized structures, marked in the
figure by dots, and under the imposition of $8$ (evenly 
spaced) radial shell constraints this nonphysical feature all but 
vanishes. 

Stemming from the need to properly describe sub-wavelength 
field characteristics, local constraints are also generally required to 
formulate relevant limits on near field phenomena.
Following Fig.~\ref{fig7_localConstraints}\textbf{c}---bounds on the 
maximization of radiative Purcell enhancement for a dipolar current source
separated from an arbitrary device by a distance $d$, again contained 
within a spherical ball of radius $R$---when only 
global constraints are considered enhancement is seen to scale 
$\propto 1/\Im\chi$ when $d\ll\lambda$ (light lines), as is 
characteristic of material loss limited resonant 
response~\cite{miller2016fundamental}. 
Upon inclusion of localized constraints defined over
concentric spherical shells (dark lines), the maximal radiation is
observed to drop by several orders of magnitude and no longer grows
with decreasing $\Im\chi$, confirming the challenge of achieving
optimal polarization fields capable of resonantly scattering
evanescent fields into propagating waves. Conversely, even with many
more localizing shells (see Ref.~\cite{molesky2020hierarchical} for
details), limits on larger domains and stronger dielectrics 
show no such saturation and exhibit $\propto d^{-3}$ diverging 
growth---implying that arbitrarily high angular momentum states can be
out-coupled with nearly fixed efficiency. Because of the resolution
arguments stated above, if additional non-symmetric localized
constraints were to be imposed, particularly in the region of the
design volume nearest the dipole, tighter limits may be anticipated.

\begin{figure*}[t!]
\centering 
  \includegraphics[width=1.0\textwidth]{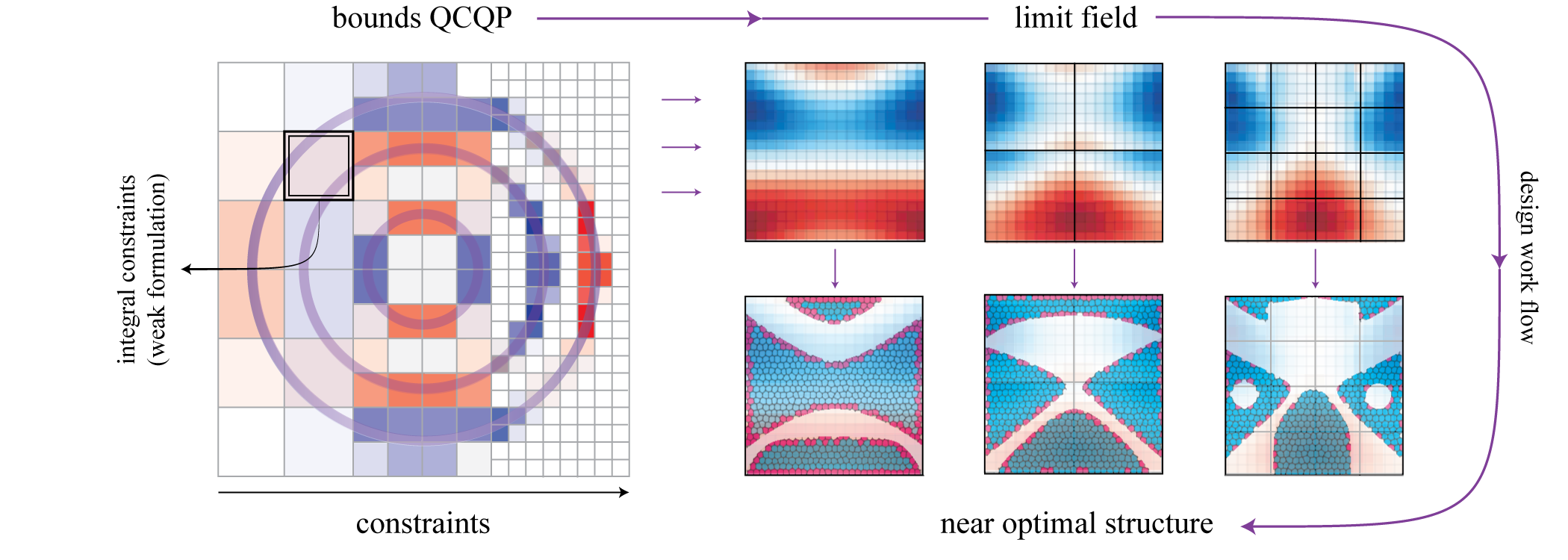} \vspace{-10
  pt} \caption{\textbf{From performance bounds to structural optimization.} 
  Through the common denominator of spatial divisions, a complementary 
  relationship exists between inverse design and performance bounds. 
  In inverse design, the non-convex optimization problem of structural 
  optimization is heuristically approximated by particular designs 
  representing local optima, with resulting device performance generally 
  increasing with additional structural degrees of freedom.
  In optimization bounds, a coarse grained version of the same 
  optimization problem is bounded by exploiting some convex
  relaxation in order to determine a value that must be
  respected by all allowed designs, with resulting limits generally 
  becoming tighter under the use of additional constraints. 
  Consequently, as the number and resolution of the local 
  constraints going into a bounds QCQP is increased, its solution 
  fields typically become increasingly better approximations of the 
  polarization field of a globally optimal structure. 
  This observation suggests a potential, presently unexplored, 
  workflow for the computational design of photonic devices: evaluate 
  the limits of an application to a tightness wherein the approximate field 
  distribution can be extracted and used as a starting point for the 
  discovery of near optimal structures.} 
  \label{fig8_outLook}
\end{figure*}
\section*{Outlook}
As exemplified by the preceding discussion and representative
examples, the nascent development of a methodology combining physical
constraints with optimization theory has already proven to be a
remarkably versatile and fruitful framework for understanding and
computing electromagnetic limits. 
Still, a panorama of challenges and opportunities remain. 
First, the core ideas behind the method are by no means
restricted to phenomena governed by Maxwell’s equations, and have
direct analogues in other domains of wave physics, such as acoustics 
and quantum mechanics. 
And even more broadly, the general concept of obtaining 
performance limits by rigorously bounding a simpler problem through relaxed 
physics is applicable to essentially any field of engineering. 
(See Ref.~\cite{zhang2021conservationlawbased} for recent applications to
quantum control.) Second, as it currently stands, the framework requires 
physical objectives to be formulated as quadratic functions of the
polarization fields, which excludes various photonic problems,
including nonlinear processes (such as the Kerr effect) or objectives
which may be more naturally expressed as eigenproblems (such as the
maximization of bandgaps~\cite{men2014robust} or the engineering of
Dirac points~\cite{lin2018Dirac}. Yet another avenue for further
research, particularly relevant to device applications, is the
incorporation of fabrication constraints. Presently, the limit
framework considers any possible structure that may fit within the
design region without regard for existing nanofabrication constraints
such as minimum feature sizes and material connectivity.

Moreover, it is not yet understood whether, or under what conditions, 
the convex relaxation techniques that are used to obtain optimization 
limits can be guaranteed to solve their associated QCQPs, i.e., whether 
the limit is equal to the true QCQP solution or just some larger
(resp. smaller) value. 
Non-affine equality constraints such as the conservation of real and 
reactive power are non-convex, and QCQPs with non-convex constraints 
are not generally thought to be solvable by any convex 
relaxation~\cite{aaronson2005guest,angeris2021heuristic}.
So far, however, a vast majority of investigations have found 
that calculated limit fields are in fact optimal solutions of the 
initial optimization problem statement~\cite{gustafsson2020upper,
molesky2020fundamental,molesky2020hierarchical},
with inclusion of large numbers of local constraints only resulting in
numerical ill-conditioning. 
Such guarantees are not merely a theoretical exercise.
So long as the underlying QCQP is actually solved, through the addition of 
increasingly finer localized constraints progressively better approximations 
to an optimal realizable polarization field emerge from bounds computations, 
and this information could be leveraged to great effect in inverse design. 
For instance, it could be used as a starting point for
adjoint optimization to recover a near optimal structure, see
Fig.~\ref{fig8_outLook}, or as a guide for the design parameters that 
should be considered to possibly realize improved performance 
characteristics. 
Furthermore, it is likely not necessary to
enforce local constraints down to the computational pixel level to
make use of these potentially powerful connections; a coarse distribution may be
enough to extract an approximate optimal structure. 
Indeed, the onset of ill-conditioning with finer local constraints suggests that,
depending on the design problem, there is a characteristic length
scale beyond which more detailed structuring becomes
unnecessary. Relatedly, while the bounds computation is convex,
it is not necessarily easy to solve numerically.  Besides
ill-conditioning resulting from the imposition of large numbers of
local constraints, numerical instabilities also occur in systems with
low loss (e.g., semi-transparent media). To address this, it may be
possible to formulate alternate constraints better suited to handle
low-loss/lossless design problems; one example is given by
Ref.~\cite{trivedi2020bounds}, though the proposed constraints appear
to provide non-trivial limits only for small design regions and low
dielectric contrasts.

The large scale of most practical photonics problems also poses a
challenge: current demonstrations of the framework are restricted to
2D \cite{trivedi2020bounds,kuang2020computational,angeris2021heuristic,molesky2021comm}
or highly symmetric domains in 3D, exploiting efficient spectral basis
representations \cite{gustafsson2020upper,molesky2020fundamental,jelinek2021fundamental}
to limit the size of the associated system matrices. Beyond these
early proof-of-concept explorations, there is much room for
development of standardized packages for evaluating limits (possibly
in conjunction with inverse methods) on 3D photonic devices by
exploiting general-purpose techniques---numerical Maxwell solvers
employing localized bases---and open-source optimization
methods~\cite{NLopt}.
\bibliography{esaLibF}
\end{document}